\documentclass[12pt,twoside]{article}
\usepackage[mathscr]{eucal}
\usepackage{amsmath,amsfonts,amssymb,amsthm,mathabx,empheq}
\usepackage{times}
\usepackage{pdfsync}
\usepackage{cite}
\usepackage{url}
\usepackage{hyperref}
\usepackage{tensor}
\usepackage{color}
\usepackage{multicol}
\usepackage{bbold}

\advance\textheight\topskip
\allowdisplaybreaks[1]

\newcommand\td{\text{d}}
\newcommand\cO{{\cal O}}
\newcommand\cM{{\cal M}}
\newcommand{\p}{\partial}

\newcommand{\be}{\begin{equation}}
\newcommand{\ee}{\end{equation}}
\newcommand{\bea}{\begin{eqnarray}}
\newcommand{\eea}{\end{eqnarray}}

\def\bz{\bar z}

\def\half{\frac12}

\def\bY{\bar Y}

\def\bp{\bar \partial}
\def\n{\nabla}
\def\bm{\bar{m}}
\def\bP{\bar P}
\def\bL{\bar{L}}

\def\bL{\bar{L}}

\newcommand{\nn}{\nonumber}
\newcommand*\xbar[1]{%
  \hbox{%
    \vbox{%
      \hrule height 0.5pt % The actual bar
      \kern0.3ex%         % Distance between bar and symbol
      \hbox{%
        \kern-0.0em%      % Shortening on the left side
        \ensuremath{#1}%
        \kern-0.0em%      % Shortening on the right side
      }%
    }%
  }%
}

\allowdisplaybreaks[1]

\DeclareFontFamily{OT1}{rsfs}{} \DeclareFontShape{OT1}{rsfs}{m}{n}{
<-7> rsfs5 <7-10> rsfs7 <10-> rsfs10}{}
\DeclareMathAlphabet{\mycal}{OT1}{rsfs}{m}{n}

\hypersetup{
colorlinks=true,
linktoc=page,    %set to all if you want both sections and subsections linked
linkcolor=blue,
citecolor=blue,
urlcolor=blue,%magenta,
colorlinks=true,
}

\def\oneone{\rlap 1\mkern4mu{\rm l}}
\def\cL{{\cal L}}

\begin{document}

\title{Four-dimensional Stationary Algebraically Special Solutions, Weyl Invariants, and Soft Hairs Beyond Large Gauge Transformations}

\author{H. L\"{u} and Pujian Mao}
\date{}

\def\mytitle{Four-dimensional Stationary Algebraically Special Solutions, Weyl Invariants, and Soft Hairs Beyond Large Gauge Transformations}

\addtolength{\headsep}{4pt}

\begin{centering}

  \vspace{1cm}

  \textbf{\Large{\mytitle}}

  \vspace{1cm}

  {\large H. L\"{u}$^{a,b,c}$ and Pujian Mao$^{a}$}

\vspace{.5cm}

\vspace{.5cm}
\begin{minipage}{.9\textwidth}\small \it  \begin{center}
     ${}^{a}$ Center for Joint Quantum Studies, Department of Physics,\\
     School of Science, Tianjin University, 135 Yaguan Road, Tianjin 300350, China
 \end{center}
\end{minipage}
\vspace{0.3cm}

\begin{minipage}{.9\textwidth}\small \it  \begin{center}
    ${}^{b}$ The International Joint Institute of Tianjin University, Fuzhou,\\
 Tianjin University, Tianjin 300350, China
 \end{center}
 \end{minipage}
\vspace{0.3cm}

\begin{minipage}{.9\textwidth}\small \it  \begin{center}
    ${}^{c}$ Peng Huanwu Center for Fundamental Theory,\\ Hefei, Anhui 230026, China
 \end{center}
 \end{minipage}

\end{centering}

\begin{center}
Emails: mrhonglu@gmail.com,\,\,\, pjmao@tju.edu.cn
\end{center}

\begin{center}
\begin{minipage}{.9\textwidth}
\textsc{Abstract}: We revisit the Ricci-flat metrics in four dimensions that are stationary and algebraically special, together with the locally asymptotically flat conditions in the generalized Bondi-Sachs framework. We show that the Einstein equation is reduced to Laplacian equation on the celestial sphere. The solutions consist of two pairs of arbitrary holomorphic and antiholomorphic functions analogous to the Virasoro modes. We prove that the higher modes of one pair of the (anti-)holomorphic function contain an infinite tower of soft hairs from the perspectives of the asymptotic supertranslation charges. We verify that different modes of the soft hairs are distinct solutions which cannot be connected by diffeomorphism, using the Weyl invariants associated to the solutions. Extending the construction to Einstein–Maxwell theory introduces a third pair of (anti-)holomorphic functions, arising from the Maxwell tensor which generates soft electric hair. We further present an exact soft-hairy solution of Einstein–Maxwell theory with a cosmological constant, offering a potential understanding of soft hair from the AdS/CFT correspondence.

\end{minipage}
\end{center}

\thispagestyle{empty}

\newpage
\tableofcontents

\newpage
\section{Introduction}

Einstein's theory of General Relativity is one of the fundamental theories of modern physics, providing a unified description of classical gravitational interaction as a geometric property of space and time. In particular, Einstein's equation relates the curvature of spacetime to the energy and momentum of matter. For any physical theory, an essential mission is to construct and study its solution space. Black holes are the most intriguing exact solutions of Einstein's theory. Nevertheless, the existence of black holes in our universe has been confirmed by both theoretical prediction \cite{Penrose:1964wq,Hawking:1973uf} and experimental observation \cite{EventHorizonTelescope:2019dse}. A longstanding problem in black hole physics is that information appears to be destroyed when a black hole forms and subsequently evaporates \cite{Hawking:1975vcx,Hawking:1974rv,Hawking:1976ra}, giving rise to the notorious information paradox. Since this paradox challenges the fundamental principle of unitarity in both classical and quantum physics, it has remained at the forefront of research for decades.

Over the past half century, numerous efforts from various perspectives have been devoted into the resolution of the information paradox, see, e.g., reviews in \cite{Mathur:2009hf,Almheiri:2020cfm,Raju:2020smc}. Well-known proposals include the page curve as unitarity evaporation\cite{Page:1993wv}, loop quantum gravity \cite{Perez:2017cmj}, the fuzzball proposal \cite{Lunin:2001jy,Mathur:2005zp,Skenderis:2008qn}, the firewall resolution \cite{Almheiri:2012rt}, the island prescription inspired by holography \cite{Penington:2019npb,Almheiri:2019psf,Penington:2019kki,Almheiri:2019yqk,Almheiri:2019qdq}. In 2016, Hawking, Perry, Strominger (HPS) proposed that the vacuum in quantum gravity is not unique and black holes can carry soft hairs \cite{Hawking:2016msc}. The ingenious feature of the soft-hair proposal \cite{Hawking:2016msc} is that the new degrees of freedom can arise purely from the large gauge transformation, such as the Bondi–Metzner–Sachs (BMS) supertranslation \cite{Bondi:1962px,Sachs:1962wk}. Information about the details of the initial state is not permanently lost, instead, it can be stored by soft hairs without violating the no-hair conjecture \cite{Israel:1967wq,Carter:1971zc,Robinson:1975bv,Hawking:1971vc,Chrusciel:2012jk}. Subsequently, soft hairy solutions have been constructed by applying large gauge transformation to exact solutions \cite{Compere:2016jwb,Compere:2016hzt,Hawking:2016sgy,Haco:2018ske,Choi:2019fuq,Mao:2024pqz}. However, this type of soft hair turns on only soft modes \cite{Strominger:2013jfa,He:2014laa,Strominger:2014pwa}, for reviews, see \cite{Strominger:2017zoo}. 

The essential step towards resolving the information paradox is the discovery of enough hairs that account for the black hole entropy. A well-known example in supersymmetric theories is the existence of large classes of supergravity solutions corresponding to ``fuzzball'', i.e., horizonless bound states or microstate geometries
\cite{Mathur:2003hj,Bena:2005va,Berglund:2005vb,Bena:2006kb,Bena:2007qc,Bena:2007qc,deBoer:2009un,Lunin:2012gp,Giusto:2013bda,Bena:2015bea,Bena:2016agb,Bena:2016ypk}. Yet, the soft-hair resolution has not fully succeeded as it has been shown that the soft dynamics decomposes into superselection sectors that cannot interact with hard dynamics \cite{Mirbabayi:2016axw,Gabai:2016kuf,Gomez:2017soz,Hamada:2017uot,Bousso:2017dny,Bousso:2017rsx}. Furthermore, soft-hairy solutions are physically indistinguishable in semi-classical effect, such as the Unruh effect and Hawking radiation \cite{Compere:2019rof,Lin:2020gva}. Specifically, the main point of those criticisms is that the existence of new symmetries does not necessarily imply the existence of new black hole dynamics.

The purpose of this paper is to specify and investigate in detail the new black hole dynamics associated with soft hairs beyond large gauge transformations. In addition to providing exact solutions with soft hairs, a crucial aspect of this work is the systematic construction of a method to distinguish these soft-hairy solutions from local degrees of freedom. This point is particularly challenging as an independent problem in General Relativity, because diffeomorphism covariance implies that the same solution can appear in completely different forms in different coordinate systems. 

We study the general four-dimensional vacuum solutions of the Einstein equation with respect to the following conditions. We assume that the spacetime is algebraically special, locally asymptotically flat in the generalized Bondi-Sachs framework \cite{Bondi:1962px,Sachs:1962wk} including the twist function \cite{Adamo:2009vu,Mao:2024jpt}, and stationary, that is, there exists a Killing vector that is timelike, at least near null infinity.\footnote{For locally asymptotically flat spacetime,
  there can be Misner-type strings~\cite{Misner:1967} at one or more locations
  on the celestial sphere \cite{Barnich:2009se,Barnich:2010ojg,Barnich:2010eb}
  which allows one to include the Taub-NUT solution
  \cite{Kol:2019nkc,Godazgar:2019dkh}.} Moreover, we set the boundary to be a unit sphere with punctures. We show that the Einstein equation is reduced to a linear second order differential equation on the celestial sphere. The solution space contains two pairs of arbitrary functions, $(L,\bL)$ and $(f,\bar f)$, where $L$ and $f$ are holomorphic, and $\bL$ and $\bar f$ are anti-holomorphic on the celestial sphere in the stereographic coordinates $(z,\bz)$. The extension to electrovacuum case introduces a third pair of (anti-)holomorphic functions $(Q,\bar Q)$, arising from the Maxwell fields. These functions are very similar to the Virasoro modes when the holomorphic and anti-holomorphic functions are expanded in convergent power series. 

We compute the asymptotically conserved charges at null infinity for the general solution space and find that the higher modes of $(L,\bL)$ and $(Q,\bar Q)$ contain nontrivial supertranslation and large $U(1)$ gauge transformation charges. However, the contribution of those higher modes to the mass and electric charge vanishes. Hence, they are exactly the soft-hairy solutions in the sense of the proposal of HPS \cite{Hawking:2016msc}. 

We develop a generic algorithm in the Newman-Penrose (NP) formalism \cite{Newman:1961qr} to compute various Weyl invariants in a highly efficient manner. We then propose using Weyl invariants to construct unique relations for each mode of the solution, in order to verify that the different modes coming from $(L,\bL)$ are independent vacuum solutions that are not related by coordinate transformations when the $(f,\bar f)$ modes are turned off. To our knowledge, these constitute the first examples of soft-hairy exact vacuum solutions that are not merely consequences of diffeomorphism. 

Moreover, we present an exact soft hairy solution of Einstein-Maxwell theory with a cosmological constant, which generalizes the Reissner–Nordstr\"{o}m-(A)dS solution by incorporating (anti)-holomorphic functions $(Q,\bar Q)$ from the Maxwell fields. Potentially, this solution offers a new avenue to explore the physical properties of soft hair via the anti-de Sitter/conformal field theory (AdS/CFT) correspondence.

%%%%%%%%%%%%%%%%%%%%%%%%%%%%%%%%%%%%%%%%%%%%%%%%%%%%%%%%%%%%%%%%%%%%%%%%%%%%%%%%%%%%%%%%%%%%%%%%%%%%%%%%%%%%%%%%%%%%%%%%%%%%%%%%%%%%%%%%%%%%%%%%%%%%%%%%%%%%%%%%%%%%%%%%%%%%%%%%%%%%%%%%%%%%%%%%%%%%%%%%%%%%%%%%%%%%%%%%%%%%%%%%%%%%%%%%%%%%%%%%

\section{Vaccum solution space}

The algebraically special vacuum solution exhibits a remarkable simplification in the radial direction \cite{Stephani:2003tm}. Recently, the algebraically special solution space was revisited in the NP formalism \cite{Mao:2024jpt}, incorporating the locally asymptotically flat conditions \cite{Bondi:1962px,Sachs:1962wk,Newman:1961qr,Newman:1962cia}. In this work, we impose two additional constraints: first, the stationary condition, requiring the existence of a global Killing vector; and second, that the boundary metric is a unit sphere with punctures. We adopt the $(u,r,z,\bz)$ coordinate system, where $\frac{\p}{\p r}$ points to the repeated principal null direction with $r$ being the affine parameter of the null geodesics, $(z,\bz)$ are the stereographic coordinates that are related to the usual angular variables via $z=e^{i\phi}\cot\frac{\theta}{2}$. Any field of the solution space is $u$-independent, i.e., $\frac{\p}{\p u}$ is a Killing vector. The future null infinity is precisely the submanifold $r=\infty$ with the topology $\mathbb{R}\times \mathbb{C}_*$. The boundary line element is $\td s^2=\frac{4}{(1+z\bz)^2}\td z \td\bz$. The solutions with respect to the above conditions were originally obtained in \cite{Trim}, see also chapter 29.2.5 of \cite{Stephani:2003tm}. Here, we will present a new organization of the equation of motion in the NP formalism, which could benefit the classification of the solutions.

Let $P_S=\frac{1+z\bz}{\sqrt2}$, the stationary algebraically special solution space is described by a co-tetrad system \cite{Mao:2024jpt}
\begin{align}
&l=\td u - \frac{\xbar \cM}{P_S} \td z - \frac{\cM}{P_S} \td \bz,\nn\\
&n= \td r  - \frac{\xbar\omega^0 }{P_S} \td z  -  \frac{\omega^0 }{P_S} \td \bz + \left[ \frac12 + \frac{rK+J\Sigma}{r^2+\Sigma^2} \right] l ,\label{tetrad}\\
&m=-\frac{1}{P_S} \left(r + i\Sigma \right) \td z ,\quad\quad \bm= -\frac{1}{P_S} \left(r - i\Sigma \right) \td \bz ,\nn
\end{align}
where $\cM$ is an arbitrary complex function of variables $(z,\bz)$ and $\xbar \cM$ is its complex conjugate. $K$ and $J$ are two real functions which are the real and imaginary parts of the arbitrary complex function $\Psi_2^0$ of variables $(z,\bz)$,
\be
\begin{split}
&K=\frac12\left[\Psi_2^0(z,\bz) + \xbar\Psi_2^0(z,\bz)\right],\\
&J=-\frac{i}{2}\left[\Psi_2^0(z,\bz) - \xbar\Psi_2^0(z,\bz)\right].
\end{split}
\ee
$\Sigma$ is a real function indicating the twist of the geodesic congruence of $l$ which is fixed by $\cM$ from
\be
\Sigma=-\frac{i P_S^2}{2}\left[ \bp (\frac{\xbar \cM}{P_S} )- \p (\frac{  \cM}{P_S}) \right]. 
\ee
The twist function $\Sigma$ determines $\omega^0$ and $\xbar\omega^0$ as $\omega^0=- i P_S \bp \Sigma, \,\,\xbar\omega^0= i P_S \p \Sigma$. We use $\p$ and $\bp$ to denote $\p_z$ and $\p_{\bz}$ for notational brevity. The unknown complex functions $\cM$, $\Psi_2^0$ and their complex conjugate are constrained by \cite{Mao:2024jpt}
\be
\Psi_2^0-\xbar\Psi_2^0=  -2 i (\Sigma + P^2_S \p\bp \Sigma), \quad \bp \Psi_2^0=0, \quad \p \xbar\Psi_2^0=0 .
\ee
These equations comprise all the constraints from the Einstein equation and the Bianchi identities. The above construction encompasses all the algebraically special stationary solutions under locally asymptotically flat conditions. The spacetime metric in $(+,-,-,-)$ signature is obtained from the co-tetrad system as
$g_{\mu\nu}=2 n_{(\mu} l_{\nu )} - 2 m_{(\mu} {\bm}_{\nu)}$. Clearly, the condition $\bp\Psi_2^0=0$ implies that $\Psi_2^0$ is a holomorphic function which we denote as $\Psi_2^0=L(z)$. Similarly, $\xbar\Psi_2^0=\bL(\bz)$.\footnote{Note that we need to remove the south and north poles from the boundary sphere. Otherwise, $L=L_0+L_1 z+L_2 z^2$ as the globally defined solutions, because $\p \frac{1}{\bz}=\pi \delta^2(z)$ from the usual two-dimensional conformal field theory knowledge with complex variables.} Then, the twist function $\Sigma$ is determined from the differential equation
\be\label{Sigma}
\Sigma + \frac{(1+z\bz)^2}{2} \p\bp \Sigma=\frac{i}{2}(L-\bL).
\ee
Now the above equation is the only constraint left for the whole system. Each solution of \eqref{Sigma} can uniquely give an exact vacuum solution of Einstein theory. We find that it is convenient to shift the twist function $\Sigma$ by
\be
\Sigma=A + \frac{i}{2} (L-\bL).
\ee
Basically, the twist function includes two parts, the imaginary part of $\Psi_2^0$ and the intrinsic part $A$ that satisfies the second-order linear differential equation
\be\label{A}
\frac{(1+z\bz)^2}{2} \p\bp A + A=0.
\ee
The salient feature of \eqref{A} is its linearity: it is simply the Laplacian equation on a unit-round sphere with an eigenvalue $-2$. Consequently, any linear combination of two solutions is again a solution. Each solution of \eqref{A} generates a tower of exact solutions, since $L$ and $\bL$ are completely arbitrary, analogous to the Virasoro modes. Note that the reality condition requires that $L$ and $\bL$ be complex conjugates of each other, which we impose. Once $A$ is specified, the functions $\Sigma$, $\cM$ can be correspondingly determined. It is worth noting that in Kerr' original work \cite{Kerr:1963ud}, the co-tetrad system with stationary condition was identical to our \eqref{tetrad}. Kerr's strategy was to substitute $\cM$ into \eqref{Sigma} and solve for $\cM$ directly, see also \cite{Stephani:2003tm} for a pedagogical review.

A generic solution of \eqref{A} from a potential is given by
\be\label{f}
A=\frac{i}{2} P_S^2 \left[\p\left(\frac{f(z)}{P_S^2}\right) - \bp\left(\frac{\bar f (\bz)}{P_S^2}\right)\right] .
\ee
The corresponding $\cM$ for the generic solution is
\be\label{our}
\cM=\frac{\sqrt{2}}{\bz} \bL   + \frac{ f}{P_S}+P_S \bp C(z,\bz),
\ee
where $C(z,\bz)$ is trivial, denoting the supertranslation freedom associated with a shift in the time direction $u\rightarrow u-C$. Those solutions are equivalent to the ones presented in chapter 29.2.5 of \cite{Stephani:2003tm} by a shift of $f$ in \eqref{f},
\be
f\rightarrow f-z^2 \int \frac{L(z)}{z^2} \td z,
\ee
up to the supertranslation freedom. The complete solution space can be classified by the mode expansion of the holomorphic functions, 
\begin{align}
&L=\sum\limits_{n} L_n z^n=\sum\limits_{n} (M_n + i N_n)z^n,\\
&f=\sum\limits_{n} f_n z^n=\sum\limits_{n} (X_n + i Y_n)z^n,
\end{align}
where we introduce the real parameters $M_n$, $N_n$, $X_n$, and $Y_n$. For some particular modes, $(M_0,N_0)$ are related to the mass and NUT parameters, $Y_1$ is related to the rotation. 

The singularity of Ricci-flat spacetime is normally characterized by the principal invariants of the Weyl tensor, such as the Kretschmann scalar $I_1=W^{\mu\nu\alpha\beta}W_{\mu\nu\alpha\beta}$ and Chern-Pontryagin scalar $I_2=\frac12 \epsilon_{\mu\nu\alpha\beta}{W^{\alpha\beta}}_{\sigma\rho} W^{\mu\nu\sigma\rho}$, which are given in the NP formalism for algebraically special spacetime as \cite{Cherubini:2002gen}
\be
\begin{split}
I_1=24\left[(\Psi_2)^2+(\xbar\Psi_2)^2\right],\\ I_2=24i\left[(\Psi_2)^2-(\xbar\Psi_2)^2\right].
\end{split}
\ee
Recalling the solutions, we obtain
\be
\begin{split}
&I_1=\frac{24L^2(r-i\Sigma)^6+24\bL^2(r+i\Sigma)^6}{(r^2+\Sigma^2)^6},\\
&I_2=i\frac{24L^2(r-i\Sigma)^6 - 24\bL^2(r+i\Sigma)^6}{(r^2+\Sigma^2)^6},
\end{split}
\ee
which indicate that the spacetime singularity is located at $r^2+\Sigma^2=0$. In particular, the singularity should be a circle at $r=0$ parametrized by solution of $\Sigma(z,\bz)=0$ if $r$ is a real variable. 

As a closing remark, one can prove for the subclass of Petrov type-D the uniqueness of the Kerr-Taub-NUT solution from our generic set-up, which is presented in Appendix \ref{typeD}. We set the boundary a unit sphere with punctures. The results can be recovered in principle for arbitrary two-dimensional boundary surface, including in particular the plane case as demonstrated in Appendix \ref{2}.

%%%%%%%%%%%%%%%%%%%%%%%%%%%%%%%%%%%%%%%%%%%%%%%%%%%%%%%%%%%%%%%%%%%%%%%%%%%%%%%%%%%%%%%%%%%%%%%%%%%%%%%%%%%%%%%%%%%%%%%%%%%%%%%%%%%%%%%%%%%%%%%%%%%%%%%%%%%%%%%%%%%%%%%%%%%%%%%%%%%%%%%%%%%%%%%%%%%%%%%%%%%%%%%%%%%%%%%%%%%%%%%%%%%%%%%%%%%%%%%%

\section{Soft hairs of the vacuum solution} 

The asymptotic symmetries are those residual gauge transformations that leave the field configurations under consideration asymptotically invariant \cite{Bondi:1962px,Sachs:1962wk,Sachs:1962zza,Henneaux:1985tv,Brown:1986nw}. In the NP formalism, a gauge transformation consists of a combination of a diffeomorphism and a local Lorentz transformation. The transformations acting on the tetrad and the spin connection are given by 
\be
\begin{split}
\delta_{\xi,\Lambda}{e_a}^\mu = &{\xi}^\nu\partial_\nu
{e_a}^\mu-\p_\nu{\xi}^\mu{e_a}^\nu +{\Lambda_a}^b{e_b}^\mu, \\
\delta_{\xi, \Lambda} \Gamma_{a b c} = &{\xi}^\nu \partial_\nu \Gamma_{a b c} - e_c^\mu \p_\mu {\Lambda}_{a b}\\
&+ {\Lambda_a}^{d}\Gamma_{dbc}+ {\Lambda_b}^{d}\Gamma_{adc} + {\Lambda_c}^{d}\Gamma_{abd} .
\end{split}
\ee
The asymptotic symmetries of the algebraically special solution space were derived in detail in \cite{Mao:2024jpt}. After imposing the stationary condition and taking the boundary to be a punctured unit sphere, the corresponding symmetry parameters are obtained as
\be
\begin{split}
&\xi^u=T(z,\bz), \quad \xi^r=0,\\
&\xi^z=y_+(z^2+1) - i y_- (z^2-1)+i y_0 z, \\
&\xi^{\bz}=y_+(\bz^2+1) + i y_- (\bz^2-1) - i y_0 \bz,\\
&\Lambda^{12}=\Lambda^{13}=\Lambda^{14}=\Lambda^{23}=\Lambda^{24}=0,\\ 
&\Lambda^{34}=-i y_0 + y_+ (\bz - z) + i y_- (z+ \bz),
\end{split}
\ee
where $y_0$ and $y_{\pm}$ are real constant parameters representing the $SO(3)$ rotations. The asymptotic symmetry algebra consists of the semidirect sum of supertranslations, parametrized by an arbitrary function $T$ on the celestial sphere, with $SO(3)$ rotations. The Lorentz boosts present in the standard global BMS symmetries \cite{Bondi:1962px,Sachs:1962wk,Sachs:1962zza} are excluded by the stationary configuration. The transformation law on the solution space is
\be\label{Ktransf}
\begin{split}
&\delta \cM=Y \p \cM  + \bY \bp \cM + \bp \bY \cM  - P_S \bp T,\\
&\delta \Sigma=Y \p \Sigma  + \bY \bp \Sigma,\\
&\delta K = Y \p  K  + \bY \bp  K, \\
&\delta J = Y \p  J  + \bY \bp  J .
\end{split}
\ee
The boundary charges can be computed directly in the tetrad system \cite{Barnich:2019vzx,Godazgar:2020kqd}. The supertranslation charge is simply
\be
Q_T =\frac{1}{8\pi G} \int_{\partial \Sigma} \frac{\td z \td \bz}{P_S^2} T K ,
\ee
where $\partial \Sigma$ can be any constant-$u$ two-surface on the boundary. Because we are in a stationary configuration, the charge is fully conserved. Therefore, the asymptotic charge evaluated on any constant-$u$ two-surface at the boundary must yield the same result. Expanding the supertranslation parameter in spherical harmonics $T=\sum_{l,m} T_{l,m}Y_{l,m}$, the supertranslation charge for the positive $n^{\rm th}$ mode of the solution
\be
\begin{split}
K_n&=\frac12 \left[(M_n+i N_n) z^n + (M_n-i N_n)\bz^n \right]\\
&=M_n \cot^n\frac{\theta}{2} \cos(n\phi) - N_n \cot^n\frac{\theta}{2}\sin(n\phi), 
\end{split}
\ee
can be obtained as 
\be
Q_T \sim (T_{n,n}+T_{n,-n}) M_n - i (T_{n,n}-T_{n,-n}) N_n,
\ee
where $Y_{n,\pm n}\sim\sin^n\theta e^{\pm in\phi}$. Since different modes have different prefactors depending on the normalization conditions, we use the $\sim$ symbol to indicate only the essential charge content of each mode. The charge of any given mode can be completely determined by performing a surface integral on the boundary. Similarly, for the negative modes of the solution, we have
\be
Q_T \sim (T_{n,n}+T_{n,-n}) M_{-n} + i (T_{n,n}-T_{n,-n}) N_{-n}.
\ee
Specifically, we see that there is only one component $Y_{0,0}$ that defines the mass charge of the solution from the zero mode $(M_0,N_0)$.

%%%%%%%%%%%%%%%%%%%%%%%%%%%%%%%%%%%%%%%%%%%%%%%%%%%%%%%%%%%%%%%%%%%%%%%%%%%%%%%%%%%%%%%%%%%%%%%%%%%%%%%%%%%%%%%%%%%%%%%%%%%%%%%%%%%%%%%%%%%%%%%%%%%%%%%%%%%%%%%%%%%%%%%%%%%%%%%%%%%%%%%%%%%%%%%%%%%%%%%%%%%%%%%%%%%%%%%%%%%%%%%%%%%%%%%%%%%%%%%%

\section{Weyl invariants} 

The exact solutions contain non-vanishing supertranslation hairs encoded by a pair of (anti-)holomorphic functions $(L,\bL)$ which can be characterized by a mode expansion. However, a crucial feature has not yet been addressed that if different modes are independent solutions. Naively, this appears doubtful, based on lessons from lower-dimensional gravity, where the Virasoro modes in the solution space are turned on by large gauge transformations \cite{Banados:1998gg}, see also recent developments relevant to soft hairs in \cite{Afshar:2016wfy,Sheikh-Jabbari:2016unm,Sheikh-Jabbari:2016lzm,Afshar:2016uax,Afshar:2016kjj}. We will resolve this issue by examining relations among the Weyl invariants of the solutions. We will turn off the modes from $f$ and focus exclusively on the supertranslation hairs in this work.

The simplest Weyl invariant is the square of the Weyl tensor as defined previously, which is given by
\be\label{W2}
W_2=24\left[\left(\frac{L}{\varrho^3}\right)^2+\left(\frac{\bL}{\xbar\varrho^3}\right)^2\right],
\ee
where $\varrho=r+i\Sigma$ and $\Sigma=\frac{i}{2} (L-\bL)$. The next invariant of algebraic construction without any derivative is the cubic of the Weyl tensor $W_3=W^{\mu\nu\alpha\beta}W_{\mu\nu\rho\sigma}{W_{\alpha\beta}}^{\rho\sigma}$, which is given by
\be\label{W3}
W_3=48\left[\left(\frac{L}{\varrho^3}\right)^3+\left(\frac{\bL}{\xbar\varrho^3}\right)^3\right],
\ee
when the solutions are inserted. The algorithm of deriving the invariant in the NP formalism is detailed in Appendix \ref{NPreview}-\ref{twoderivative}. For Schwarzschild solution, one can verify that $(W_2)^3=12 (W_3)^2$. However, the Taub-NUT solution does not satisfy this relation, which indicates that Taub-NUT and Schwarzschild are different solutions. Our goal is to construct a unique relation among the Weyl invariants for each mode of $(L,\bL)$, allowing us to distinguish between them unambiguously.

Since the solution space is classified by mode expansion of $(L,\bL)$, the strategy to distinguish different modes is to construct Weyl invariants with varying numbers of derivatives acting on $(L,\bL)$. Because the Weyl invariant also depends explicitly on the radial variable $r$, we need to introduce a third Weyl invariant of algebraic construction. This allows the solution variables $L$, $\bL$, $r$ to be determined from those Weyl invariants. Naturally, one would expect to check $W_4$, i.e., the fourth power of the Weyl tensor. But the special configuration of the solution space ensures that any higher-power Weyl invariant cannot provide an independent relation sufficient to determine $L$, $\bL$, $r$.

Surprisingly, we found that the invariant $\n_e W_{abcd} \n^e W^{abcd}$ becomes algebraic when evaluated on the solution space, we denote this as $dW_2$. Another possible construction that involves one covariant derivative acting on the Weyl tensor is $\n_e W_{abcd} \n^a W^{ebcd}$. We prove that this latter invariant equals exactly half of the first one. For any algebraically special solution, $dW_2$ takes a remarkably simple form in the NP formalism as
\be
dW_2= 240 \left(\rho \Psi_2 \Delta \Psi_2 + \xbar\rho \xbar\Psi_2 \Delta \xbar\Psi_2- \tau \Psi_2 \xbar\delta \Psi_2  - \xbar\tau \xbar\Psi_2 \delta \xbar\Psi_2\right). 
\ee
Inserting the solutions, we obtain
\be\label{dW2}
dW_2=-360\left(\frac{L^2}{\varrho^8 }  + \frac{\bL^2}{\xbar\varrho^8} \right) \left(1 + \frac{L}{\varrho }  + \frac{\bL}{\xbar\varrho} \right).
\ee
Consider a Weyl invariant with derivatives acting on $L$ and $\bL$, such as
\begin{multline}
\n_\mu (W_2 )\n^\mu( W_2)
=
-20736\left(1+ \frac{L}{\varrho} + \frac{\bL}{\xbar\varrho} \right)\left(\frac{L^2}{\varrho^7} + \frac{\bL^2}{\xbar\varrho^7} \right)^2 \\
- 4608 \frac{L \bL P_S^2}{\varrho^7 \xbar\varrho^7} \left(\bp \bL + \frac{6i \bL \bp \Sigma}{\xbar\varrho} \right) \left(\p L - \frac{6i L \p \Sigma}{\varrho} \right).
\end{multline}
Then, for the zero mode solution $\p L=0=\bp \bL$, there will a unique relation between $\n_\mu (W_2 )\n^\mu( W_2)$, $W_2$, $W_3$, $dW_2$. More precisely, 
\be
\n_\mu (W_2 )\n^\mu( W_2)
=
-20736\left(1+ \frac{L}{\varrho} + \frac{\bL}{\xbar\varrho} \right)\left(\frac{L^2}{\varrho^7} + \frac{\bL^2}{\xbar\varrho^7} \right)^2,
\ee
for the zero mode solution. One can then eliminate $L$, $\bL$, $r$ using $W_2$, $W_3$, $dW_2$ to reveal the explicit relation among those four Weyl invariants. However, for the first mode, or other higher mode solution, the relation among these quantities differs from that of the zero-mode solution. This difference arises because, for higher modes with $\p L\neq0$ and $\bp \bL\neq0$, additional terms appear in $\n_\mu (W_2 )\n^\mu( W_2)$. Consequently, any higher-mode solution must not be the same solution as the zero-mode solution, as they satisfy distinct diffeomorphism-invariant relations among the Weyl invariants.

One can continue this algorithm by introducing a second Weyl invariant that involves a single derivative on $L$ and $\bL$, for example, $\n_\mu (dW_2)\n^\mu (dW_2)$, as well as a Weyl invariant containing two derivatives on $L$ and $\bL$, such as $\n_\mu (\n_\nu W_2 \n^\nu W_2 )\n^\mu(W_2)$. The precise forms of these invariants are provided in Appendix \ref{twoderivative}. Using these invariants, one can derive a unique relation between $\n_\mu (\n_\nu W_2 \n^\nu W_2 )\n^\mu(W_2)$ and the set $\{W_2$, $W_3$, $dW_2$, $\n_\mu (W_2 )\n^\mu( W_2)$, $\n_\mu (dW_2)\n^\mu (dW_2)\}$ by eliminating $L$, $\bL$, $r$, $\p L$, and $\bp \bL$ for the first mode solution. Higher-mode solutions, however, do not satisfy this relation because $\n_\mu (\n_\nu W_2 \n^\nu W_2 )\n^\mu(W_2)$ contains terms with second-order derivatives of $L$ and $\bL$.

In principle, one can consider $W_2$ and $dW_2$ as the fundamental building blocks, and by acting with $n$ covariant derivatives, one can construct a Weyl invariant with $n$ derivatives acting on $L$ and $\bL$. This procedure distinguishes the $n$-th mode solutions from all lower-mode solutions. This completes our justification that each mode of $(L,\bL)$ represents an independent solution and cannot be related to any other mode via a diffeomorphism.

Before closing this section, we comment on the role of function $f$ in some particular cases. Since $f$ is the potential for the $A$ function, different modes of $f$ do not necessarily correspond to distinct solutions. A simple example is that the real part of the first mode is trivial, yielding $A=0$. The imaginary part of the first mode corresponds to the rotation parameter when $L$ and $\bL$ take the zero mode, which is nothing but the Kerr-Taub-NUT solution. An $SO(3)$ rotation of the Kerr-Taub-NUT solution turns on the zero mode of $f$, reflecting the uniqueness of the Kerr-Taub-NUT solution of type-D. 

%%%%%%%%%%%%%%%%%%%%%%%%%%%%%%%%%%%%%%%%%%%%%%%%%%%%%%%%%%%%%%%%%%%%%%%%%%%%%%%%%%%%%%%%%%%%%%%%%%%%%%%%%%%%%%%%%%%%%%%%%%%%%%%%%%%%%%%%%%%%%%%%%%%%%%%%%%%%%%%%%%%%%%%%%%%%%%%%%%%%%%%%%%%%%%%%%%%%%%%%%%%%%%%%%%%%%%%%%%%%%%%%%%%%%%%%%%%%%%%%

\section{Extension to Einstein-Maxwell theory}
The investigation can be extended to Einstein-Maxwell theory in a straightforward way. The solution can be found in, e.g., chapter 30 of \cite{Stephani:2003tm}. A self-contained derivation in the NP formalism is presented in Appendix \ref{NPMaxwell}-\ref{ASS}. The full solution space is given by
\be
\begin{split}
&l=\td u - \frac{\xbar \cM}{P_S} \td z - \frac{\cM}{P_S} \td \bz, \\
&n=\left[ \frac12 + \frac{rK+J\Sigma +   Q(z) \xbar Q(\bz)}{r^2+\Sigma^2} \right] l + \td r  - \frac{\xbar\omega^0 }{P_S} \td z  -  \frac{\omega^0 }{P_S} \td \bz,\\
&m=-\frac{1}{P_S} \left(r + i\Sigma \right) \td z ,\quad\quad \bm= -\frac{1}{P_S} \left(r - i\Sigma \right) \td \bz ,
\end{split}
\ee
where $\cM$ is present in \eqref{our}, and the Maxwell scalars are given by
\be
\begin{split}
&\phi_1=\frac{Q}{(r + i\Sigma)^2},\quad  \phi_2= - \frac{P_S \p Q}{(r + i\Sigma)^2} + \frac{2 i P_S Q \p \Sigma }{(r + i\Sigma)^3}.
\end{split}
\ee
In the solution space, a third pair of holomorphic and anti-holomorphic functions, $(Q,\xbar Q)$, arises, which can be characterized by mode expansion $Q=\sum\limits_{n} Q_n z^n=\sum\limits_{n} (q_n + i p_n)z^n$. Each mode of $(Q,\xbar Q)$ corresponds to a distinct solution, as can be expected from invariants constructed from the Maxwell tensor, analogous to the Weyl invariants discussed in the previous section. The zero modes $(q_0,p_0)$ are directly related to the electric and magnetic charges. In addition to diffeomorphism, Einstein-Maxwell theory possesses a $U(1)$ gauge symmetry. The residual large $U(1)$ gauge transformations that preserves the asymptotic form of the gauge potential are parameterized by an arbitrary scalar function on the celestial sphere \cite{Strominger:2013lka,He:2014cra}, inducing the transformations $\delta_\varepsilon A_\mu=\p_\mu \varepsilon(z,\bz)$ . The associated asymptotic charges are 
\be
Q_\Lambda=\int_{\partial \Sigma} \frac{\td z \td \bz}{P_S^2} \varepsilon (Q+\xbar Q).
\ee
One can specify the soft electric hairs by studying the mode expansion of the charge in exactly the same way as the supertranslation charge.

%%%%%%%%%%%%%%%%%%%%%%%%%%%%%%%%%%%%%%%%%%%%%%%%%%%%%%%%%%%%%%%%%%%%%%%%%%%%%%%%%%%%%%%%%%%%%%%%%%%%%%%%%%%%%%%%%%%%%%%%%%%%%%%%%%%%%%%%%%%%%%%%%%%%%%%%%%%%%%%%%%%%%%%%%%%%%%%%%%%%%%%%%%%%%%%%%%%%%%%%%%%%%%%%%%%%%%%%%%%%%%%%%%%%%%%%%%%%%%%%

\section{Soft hairy solution with a cosmological constant} 

Remarkably, a subclass of the soft hairy solution of Einstein-Maxwell theory can be extended to include a cosmological constant according to the theorem in chapter 28.4 of \cite{Stephani:2003tm}. The line-element and the Maxwell tensor satisfying the equation of motion of the $\Lambda$-Einstein-Maxwell theory,
\be
G_{\mu\nu} - \frac{3}{\ell^2} g_{\mu\nu} =  \frac12\left(F_{\alpha \mu} {F^\alpha}_\nu - \frac14 F^2 g_{\mu\nu}\right),
\ee
are given by
\be
\begin{split}
&\td s^2 = \bigg(- \frac{ r^2}{\ell^2} - 1 + \frac{2M}{r} - \frac{Q(z) \xbar Q(\bz) }{r^2} \bigg) \td u^2 \\
&\hspace{1.5cm}- 2 \td u \td r + \frac{2r^2}{P_S^2} \td z \td \bz, \qquad F=\td A,\\
&A=-\frac{Q+\xbar Q}{r} \td u + \frac{2Q}{z(1+z \bz)} \td z + \frac{2\xbar Q}{\bz (1+z \bz)} \td \bz.
\end{split}
\ee
By considering a negative cosmological constant, this solution potentially opens a new avenue for investigating soft hair in the context of the AdS/CFT correspondence. We leave these investigations for future work.

\section{Conclusions} 

In this paper, we construct general solutions under the assumptions of stationary algebraically special and locally asymptotically flat conditions. The Einstein equations can be solved completely in terms of two pairs of arbitrary holonomic and anti-holonomic functions. We show that the higher modes in $(L, \bar L)$ correspond to soft-hairy solutions as they carry nontrivial supertranslation charges. We verify that different modes of $(L, \bar L)$ represent distinct solutions by examining the relations among Weyl invariants. The system can be naturally extended to Einstein-Maxwell theory, where a third pair of  holonomic and anti-holonomic functions emerges from the Maxwell field, giving rise to soft electric hair. Finally, we provide an intriguing soft hairy solution with a cosmological constant, which may provide a holographic perspective on soft hair.

%%%%%%%%%%%%%%%%%%%%%%%%%%%%%%%%%%%%%%%%%%%%%%%%%%%%%%%%%%%%%%%%%%%%%%%%%%%%%%%%%%%%%%%%%%%%%%%%%%%%%%%%%%%%%%%%%%%%%%%%%%%%%%%%%%%%%%%%%%%%%%%%%%%%%%%%%%%%%%%%%%%%%%%%%%%%%%%%%%%%%%%%%%%%%%%%%%%%%%%%%%%%%%%%%%%%%%%%%%%%%%%%%%%%%%%%%%%%%%%%

\section*{Acknowledgments} 

P.M.~would like to thank Peng Cheng, Zhengwen Liu, Xiaoning Wu, Runqiu Yang, Hongbao Zhang, and Weicheng Zhao for useful discussions. This work is supported in part by the National Natural Science Foundation of China (NSFC) under Grant  No.~11935009 and by Tianjin University Self-Innovation Fund Extreme Basic Research Project Grant No. 2025XJ21-0007. H.L.~is also supported in part by NSFC Grants No.~12375052 and No.~12247103. P.M.~is also supported in part by NSFC Grant No.~12475059.

\appendix

\section{Uniqueness of type-D solution}
\label{typeD}

Having obtained the solution space of our general ansatz, we now consider the type-D subclass, since important metrics such as the Kerr metric belongs to type D. In our initial configuration for the tetrad system, the basis vector $l$ is a repeated principal null direction of the Weyl tensor. Type-D metrics are those admitting another repeated principal null direction. From the standard analysis of algebraic classification in the NP formalism \cite{Chandrasekhar}, the type-D condition is that the two roots of the quadratic equation $\Psi_4+4\bar x \Psi_3+6\bar x^2\Psi_2=0$ for $\bar x$ coincide. Then the other repeated principal null direction is simply $n + x\bm + \bar x m + x\bar x l$. The coincidence of the two roots implies that $2(\Psi_3)^2=3\Psi_2 \Psi_4$, which is equivalent to
\be\label{typeDcondition}
L\p(P_S^2\p L)=\frac43 P_S^2 (\p L)^2,\,\, L \p(P_S^2 \p \Sigma)=P_S^2 \p L \p \Sigma.
\ee
The first equation can be written as
\be
\frac{\bz}{1+z \bz} (L\p L) =\frac23 (\p L)^2 - \frac12 L \p^2 L.
\ee
Recalling that $L$ is a holomorphic function of $z$, the only solution to the above equation is $\p L=0$, implying that $L$ should contain only its zero mode. Setting $L$ as a constant, the second equation in \eqref{typeDcondition} is reduced to 
\be
\p(P_S^2 \p A)=0.
\ee
Note also the complex conjugate of this equation
\be
\bp(P_S^2 \bp A)=0.
\ee
The generic solution to the above two equations is
\be
A=\frac{a_0(z\bz-1)+ a_+ (z+\bz) - i a_- (z -\bz) }{1 + z\bz} + A_0,
\ee
where $a_0$, $a_\pm$, and $A_0$ are four real parameters. Combined with the constraint as solution of Einstein equation for the function $A$, one obtains $A_0=0$. Thus, we have obtained the complete set of type-D solutions in our setup. The generic type-D solution is given by
\begin{align}
&l=\td u - \frac{\xbar \cM}{P_S} \td z - \frac{\cM}{P_S} \td \bz,\nn\\
& n=\left[ \frac12 + \frac{r M_0 + N_0 \Sigma}{r^2+\Sigma^2} \right] l + \td r  - \frac{\xbar\omega^0 }{P_S} \td z  -  \frac{\omega^0 }{P_S} \td \bz,\label{D}\\
&m=-\frac{1}{P_S} \left(r + i\Sigma \right) \td z ,\quad\quad \bm= -\frac{1}{P_S} \left(r - i\Sigma \right) \td \bz ,\nn
\end{align}
where $M_0$, $N_0$, $a_0$, $a_+$, and $a_-$ are five real constant parameters and
\be
\begin{split}
&\cM=\sqrt2 i N_0 z + \frac{ia_0z+ia_+-a_-}{P_S},\\
& \Sigma=\frac{a_0(z\bz-1)+ a_+ (z+\bz) - i a_- (z -\bz) }{1 + z\bz}  - N_0,\\
& \omega^0= \frac{\frac12 a_- (1+z^2) - \frac12 i a_+ (1-z^2) -i a_0 z}{P_S} .
\end{split}
\ee
We have applied the supertranslation freedom to modify the generic definition of $\cM$ where 
\be
C=2 M_0 \log\frac{1+z\bz}{z\bz}+2i N_0 \log\frac{\bz}{z}.
\ee
In particular, the Kerr solution in the Kerr-Schild coordinates corresponds to setting $a_\pm=0=N_0$.

It should be remarked that the Laplacian equation on a unit-round sphere can be solved from spherical harmonics for square-integrable solutions. Since we have
\be
\nabla^2 Y_{l,m}=-l (l+1) Y_{l,m},
\ee
the generic solution of 
\be
\frac{(1+z\bz)^2}{2} \p\bp A + A=0
\ee
satisfies an algebraic equation
\be
-l (l+1)+2=0,
\ee
which leads to $l=1$. If we further require that $L$ and $\bL$ are also square-integrable functions, the type-D solution \eqref{D} forms the complete set of square-integrable solutions.

\subsection{Transformations to Kerr-Taub-NUT solution}

Starting from the five-parameter $(M_0,N_0,a_+,a_0,a_-)$ solution, we can make the following coordinate redefinition
\be
z\to \frac{\alpha  z+ \bar{\beta }}{\beta  z-\bar{\alpha }}\,,\quad \bar{z}\to \frac{\bar{\alpha } \bar{z}+\beta }{\bar{\beta } \bar{z}-\alpha }\,,
\ee
with $ \alpha  \bar{\alpha }+\beta  \bar{\beta }=1$. The metric is invariant under a further shift of the $u$ coordinate,
\begin{multline}
 u\rightarrow u + 2i N_0 \log \frac{\alpha -\bar{\beta } \bar{z}}{\bar{\alpha }-\beta  z} 
 + \frac{2}{1 + z\bar z} \Big[i a_0 \left(\alpha  \beta  z-\bar{\alpha } \bar{\beta } \bar{z}\right)  \\
  +(a_- - i a_+) \beta\left(\bar{\beta } \bar{z}-\alpha\right) 
  +(a_- + i a_+) \bar{\beta } \left(\beta  z-\bar{\alpha }\right)\Big],
\end{multline}
with the redefinitions of the parameters $(M_0,N_0,\tilde a_+,\tilde a_0,\tilde a_-)$, where
\be\label{so3}
\left(\begin{array}{c}
a_+\\
a_0\\
a_-
\end{array}\right)
= {\mathbb M}
\left(\begin{array}{c}
\tilde a_+\\
\tilde a_0\\
\tilde a_-
\end{array}\right)\,,
\ee
and ${\mathbb M}$ is an $SO(3)$ matrix,
%\begin{widetext}
\be
{\mathbb M}=\left(
\begin{array}{ccc}
 \frac{1}{2} \left(-\alpha ^2+\beta ^2-\bar{\alpha }^2+\bar{\beta }^2\right) & \beta  \bar{\alpha }+\alpha  \bar{\beta } & -\frac{1}{2} i \left(\alpha ^2-\beta ^2-\bar{\alpha }^2+\bar{\beta }^2\right) \\
 \alpha  \beta +\bar{\alpha } \bar{\beta } & \alpha  \bar{\alpha }-\beta  \bar{\beta } & i \left(\alpha  \beta -\bar{\alpha } \bar{\beta }\right) \\
 \frac{1}{2} i \left(\alpha ^2+\beta ^2-\bar{\alpha }^2-\bar{\beta }^2\right) & i \left(\beta  \bar{\alpha }-\alpha  \bar{\beta }\right) & \frac{1}{2} \left(-\alpha ^2-\beta ^2-\bar{\alpha }^2-\bar{\beta }^2\right)\\
\end{array}\right),
\ee
%\end{widetext}
satisfying ${\mathbb M}^T {\mathbb M}=\oneone$.  
A better way to specify the relations in \eqref{so3} is given by the Cayley-Klein parameters, namely
\be
\left(
\begin{array}{cc}
a_0 & a_+ - i a_-\\
a_++ i a_- & -a_0
\end{array}\right) 
={\mathbb A}^\dagger\left(\begin{array}{cc}
\tilde a_0 & \tilde a_+ - i \tilde a_-\\
\tilde a_++ i \tilde a_- & -\tilde a_0
\end{array}\right){\mathbb A}\,,
\ee
where $\mathbb A$ is a unitary matrix, given by
\be
{\mathbb A}=i\left(
\begin{array}{cc}
 \alpha  & \beta  \\
 \bar{\beta } & -\bar{\alpha } \\
\end{array}
\right).
\ee
Clearly, the transformations yielding $\tilde a_\pm=0$ connect the five-parameter solution \eqref{D} to the Kerr-Taub-NUT solution in the Kerr-Schild coordinates. Correspondingly,
\be
\begin{split}
 a_0 = \tilde a_0 (\alpha \bar \alpha - \beta \bar \beta),\quad a_+= \tilde a_0( \bar \alpha \beta + \alpha \bar \beta),\quad
 a_-= i \tilde a_0 ( \bar \alpha \beta - \alpha \bar \beta).
\end{split}
\ee
Hence, we have shown that Kerr-Taub-NUT solutions are the unique type-D subclass within our general setup.

\section{Stationary case with arbitrary 2-surface boundary}
\label{2}
For stationary configuration of the algebraically special solution space, the co-tetrads are
\begin{align}
&l=\td u - \frac{\xbar \cM}{\bP} \td z - \frac{\cM}{P} \td\bz, \nn \\
& n=\left[ - \mu^0 + \frac{rK+J\Sigma}{r^2+\Sigma^2} \right] l + \td r  - \frac{i\xbar\eth \Sigma }{\bP} \td z  +  \frac{i\eth \Sigma }{P} \td\bz,\\
&m= -\frac{1}{\bP}\left(r+i\Sigma \right) \td z,\quad\quad \bm= -\frac{1}{P} \left( r-i\Sigma\right)  \td\bz ,\nn\\
&\Sigma=\frac{i}{2} ( \xbar\eth \cM - \eth \xbar \cM ),\qquad K=\frac12(\Psi_2^0 + \xbar\Psi_2^0),\qquad J=-\frac{i}{2}(\Psi_2^0 - \xbar\Psi_2^0),\nn
\end{align}
and the NP variables as given by
\begin{align}
&\rho=-\frac{1}{\varrho},\quad \varrho=r+i\Sigma,\quad \alpha=\frac{\alpha^0}{\varrho},\quad \alpha^0=\frac12 \bP \p \ln P, \qquad \beta=-\xbar\alpha ,\nn\\
& \tau=0=\lambda, \quad\quad   \gamma=-\frac{\Psi_2^0}{2\varrho^2},\qquad \mu=\frac{\mu_0}{2\xbar\varrho}-\frac{r\Psi_2^0}{\varrho^2\xbar\varrho}, \quad \quad \mu^0=-\eth\alpha^0-\xbar\eth \xbar\alpha^0, \nn \\
& \nu = \frac{\xbar\eth \Psi_2^0}{2\varrho^2} - \frac{ i \Psi_2^0  \xbar\eth \Sigma}{\varrho^3}, \quad\quad \Psi_2=\frac{\Psi_2^0}{\varrho^3},  \qquad \Psi_3=\frac{\xbar\eth \mu^0}{\varrho^2} -\frac{\xbar\eth \Psi_2^0}{\varrho^3} + \frac{3 i\Psi_2^0 \xbar\eth \Sigma}{\varrho^4},  \\
& \Psi_4=-\frac{\xbar\eth^2 \mu^0}{\varrho^2}+ \frac{4i \xbar\eth\Sigma \xbar\eth \mu^0 + \xbar\eth^2 \Psi_2^0}{2\varrho^3}  - \frac{3i\xbar\eth\Psi_2^0 \xbar\eth \Sigma + i \Psi_2^0 \xbar\eth^2 \Sigma }{\varrho^4} - \frac{6\Psi_2^0  (\xbar\eth \Sigma)^2}{\varrho^5}.\nn
\end{align}
There are three constraints from the Einstein equation and Bianchi identities
\be
\begin{split}
&\Psi_2^0-\xbar\Psi_2^0= 4 i \Sigma \mu_0 - 2i \xbar\eth  \eth \Sigma,  \\
&\eth \Psi^0_2=0, \qquad \xbar\eth \xbar\Psi^0_2=0, \quad\quad  \xbar\eth\eth \mu^0 =0 .
\end{split}
\ee
Any solution of the above system will specify a Ricci-flat solution of Einstein's theory.

\subsection{Warm-up: The plane boundary}

When choosing a plane for the 2-surface boundary, namely $P=\bP=1$, we have 
\be
\begin{split}
&\alpha^0=0,\quad \mu^0=0,\quad U^0=0,\quad \Sigma=\frac{i}{2} (\p \cM -\bp \xbar \cM ), \\
&\omega^0=- i\bp \Sigma, \quad \quad  \Psi_2^0-\xbar\Psi_2^0= -2 i\p \bp \Sigma,\\
& \bp \Psi^0_2=0 ,\quad \quad \p \xbar\Psi^0_2=0 .
\end{split}
\ee
The most general solution can be derived as
\be
\begin{split}
&\Psi_2^0=\p^2 f(z),\quad \cM=\bz f(z) - 2i h(z)  + \bp C(z,\bz),\\
&\Sigma=\p h(z) + \bp \bar h(\bz) + \frac{i}{2} \big[\bz \p f(z) - z \bp \bar f(\bz) \big],
\end{split} 
\ee
where $C$ denotes the supertranslation freedom. 

The solutions of type-D subclass are those satisfying $\p \Psi_2^0=0$ and $\p^2\Sigma=0=\bp^2\Sigma$, namely
\be
\begin{split}
& \Psi_2^0=L_0,\quad\quad \xbar\Psi_2^0=\bL_0,\\
& \Sigma=\frac{i}{2} (L_0-\bL_0)z \bz + a_1 z + \bar a_1 \bz + \Sigma_0.
\end{split}
\ee
Correspondingly,
\be
f=\frac12 L_0 z^2,\quad h=\frac12 a_1 z^2 + h_1 z,\quad \Sigma_0=h_1 + \bar h_1 .
\ee
The generic type D solution is given by
\begin{align}
&l=\td u - \xbar \cM \td z - \cM \td \bz, \nn\\
& n=\left[  \frac{r M_0 + N_0 \Sigma}{r^2+\Sigma^2} \right] l + \td r  - \xbar\omega^0 \td z  -  \omega^0 \td \bz,\label{Dplane}\\
&m=- \left(r + i\Sigma \right) \td z ,\quad\quad \bm= - \left(r - i\Sigma \right) \td \bz ,\nn
\end{align}
where the imaginary part of $a_1$ can be removed from a constant shift of the $(z,\bz)$ coordinates and we denote the real part of $a_1$ as $b$,
\be
\begin{split}
&\cM=-\frac{ i }{2} N_0 \bz z^2 - i b z^2 - i \Sigma_0 z ,\\
& \Sigma= b z + b \bz + \Sigma_0 - N_0 z\bz,\\
& \omega^0= -i (b - N_0 z) .
\end{split}
\ee
We have applied the supertranslation freedom to modify the generic definition of $\cM$ where 
\be
C=-\frac14 M_0 \bz^2 z^2 + i (h_1 - \bar h_1 ) z \bz.
\ee
Now there are four real free parameters in the metric $M_0$, $N_0$, $\Sigma_0$, and $b$.

\section{NP formalism and algebraically special spacetime}
\label{NPreview}

The NP formalism \cite{Newman:1961qr} is a tetrad formalism with two real null basis, denoted as $e_1=l=e^2,\;e_2=n=e^1$, and two complex null basis $\;e_3=m=-e^4,\;e_4=\bar{m}=-e^3$. The two complex basis are conjugate of each other to guarantee that the metric constructed from the tetrad is real. The null basis vectors are set to have the following orthogonality and normalization conditions,
\be\label{tetradcondition}
\begin{split}
 l\cdot m=l\cdot\bm=n\cdot m=n\cdot\bm=0,\quad
 l\cdot n=1,\quad m\cdot\bm=-1.
\end{split}
\ee
The spin connection is labeled by 12 complex scalars denoted by Greek symbols,
%\begin{widetext}
\be\label{coefficient}
\begin{split}
&\kappa=\Gamma_{311}=l^\nu m^\mu\nabla_\nu l_\mu,\quad \pi=-\Gamma_{421}=-l^\nu \bar{m}^\mu\nabla_\nu n_\mu, \\
& \epsilon=\half(\Gamma_{211}-\Gamma_{431})=\half(l^\nu n^\mu\nabla_\nu l_\mu - l^\nu \bar{m}^\mu\nabla_\nu m_\mu),\\
&\tau=\Gamma_{312}=n^\nu m^\mu\nabla_\nu l_\mu,\quad \nu=-\Gamma_{422}=-n^\nu \bar{m}^\mu\nabla_\nu n_\mu,\\
& \gamma=\half(\Gamma_{212}-\Gamma_{432})=\half(n^\nu n^\mu\nabla_\nu l_\mu - n^\nu \bar{m}^\mu\nabla_\nu m_\mu),\\
&\sigma=\Gamma_{313}=m^\nu m^\mu\nabla_\nu l_\mu,\quad \mu=-\Gamma_{423}=-m^\nu \bar{m}^\mu\nabla_\nu n_\mu,\\
& \beta=\half(\Gamma_{213}-\Gamma_{433})=\half(m^\nu n^\mu\nabla_\nu l_\mu - m^\nu \bar{m}^\mu\nabla_\nu m_\mu),\\
&\rho=\Gamma_{314}=\bar{m}^\nu m^\mu\nabla_\nu l_\mu,\quad \lambda=-\Gamma_{424}=-\bar{m}^\nu \bar{m}^\mu\nabla_\nu n_\mu,\\
& \alpha=\half(\Gamma_{214}-\Gamma_{434})=\half(\bar{m}^\nu n^\mu\nabla_\nu l_\mu - \bar{m}^\nu \bar{m}^\mu\nabla_\nu m_\mu).
\end{split}
\ee
%\end{widetext}
The Weyl tensor is represented by five complex scalars, denoted as $\Psi_0$, $\Psi_1$, $\Psi_2$, $\Psi_3$, $\Psi_4$,
\be
\begin{split}
\Psi_0=-C_{1313},\quad \Psi_1=-C_{1213},\quad \Psi_2=-C_{1342},\quad
\Psi_3=-C_{1242},\;\;\Psi_4=-C_{2324}. 
\end{split}
\ee
We would refer to \cite{Chandrasekhar} for the other notations.

For algebraically special spacetime, by default, there is at least one repeated principal null direction (PND). If the repeated PND is selected as null basis $l$, the Goldberg-Sachs theorem \cite{Goldberg} yields that $\sigma=0=\kappa$, accompanied by the conditions of the Weyl scalars $\Psi_0=0=\Psi_1$. On top of that, one can turn off $\epsilon$ by a third class of tetrad rotation, which can not touch the conditions $\sigma=0=\kappa$ nor $\Psi_0=0=\Psi_1$. Thus, the PND $l$ is tangent to a null geodesic with affine parameter. Then, one can choose a coordinates system $(u,r,x^A)$ to arrange the tetrad as
\be
\begin{split}
& n=W\frac{\p}{\p u} + U \frac{\p}{\p r} + X^A \frac{\p}{\p x^A},\qquad 
l=\frac{\p}{\p r},\\
& m= M\frac{\p}{\p u} + \omega\frac{\p}{\p r} + L^A \frac{\p}{\p x^A}.
\end{split}
\ee
Here, we do not have any delicate arrangement for the choice of the non-radial coordinates. The basis vectors are designated by special symbols 
\begin{align}
D=l^\mu\p_\mu,\;\;\;\;\Delta=n^\mu\p_\mu,\;\;\;\;\delta=m^\mu\p_\mu,
\end{align}
as directional derivatives.

One can obtain the following relations from \eqref{tetradcondition} and \eqref{coefficient},
\be
\begin{split}\label{1}
l^\nu \n_\nu l_\mu=&(\epsilon+\bar \epsilon)l_\mu -\kappa \bm_\mu - \bar \kappa m_\mu,\\
n^\nu \n_\nu l_\mu=&(\gamma+\bar \gamma)l_\mu -\tau \bm_\mu - \bar \tau m_\mu,\\
m^\nu \n_\nu l_\mu=&(\beta+\bar \alpha)l_\mu -\sigma \bm_\mu - \bar \rho m_\mu,\\
\n_\nu l_\mu =& (\epsilon+\bar \epsilon)n_\nu l_\mu -\kappa n_\nu \bm_\mu - \bar \kappa n_\nu m_\mu + (\gamma+\bar \gamma) l_\nu l_\mu -\tau l_\nu \bm_\mu - \bar \tau l_\nu m_\mu \\
& - (\beta+\bar \alpha) \bm_\nu l_\mu +\sigma \bm_\nu \bm_\mu + \bar \rho \bm_\nu m_\mu - (\bar \beta+ \alpha) m_\nu l_\mu + \xbar \sigma m_\nu m_\mu + \rho m_\nu \bm_\mu.
\end{split}
\ee
For algebraically special case with $l$ affine parametrized, one obtains
\be\label{dL}
\begin{split}
L_{\nu\mu}\equiv &\n_\nu l_\mu = (\gamma+\bar \gamma) l_\nu l_\mu -\tau l_\nu \bm_\mu - \bar \tau l_\nu m_\mu   + \rho m_\nu \bm_\mu  \\
&+ \bar \rho \bm_\nu m_\mu
 - (\beta+\bar \alpha) \bm_\nu l_\mu - (\bar \beta+ \alpha) m_\nu l_\mu.
\end{split}
\ee
Then, the second group
\be
\begin{split}\label{n}
l^\nu \n_\nu n_\mu=&-(\epsilon+\bar \epsilon)n_\mu + \bar\pi \bm_\mu + \pi m_\mu,\\
n^\nu \n_\nu n_\mu =& -(\gamma + \xbar\gamma) n_\mu + \bar \nu \bm_\mu + \nu m_\mu,\\
m^\nu \n_\nu n_\mu = &- (\xbar\alpha + \beta)n_\mu + \xbar\lambda \bm_\mu + \mu m_\mu,\\
\n_\nu n_\mu=&-(\epsilon+\bar \epsilon) n_\nu n_\mu + \bar\pi n_\nu \bm_\mu + \pi n_\nu m_\mu   - (\gamma + \xbar\gamma) l_\nu n_\mu + \bar \nu l_\nu \bm_\mu + \nu l_\nu m_\mu\\
&  + (\xbar\alpha + \beta) \bm_\nu n_\mu - \xbar\lambda \bm_\nu \bm_\mu - \mu \bm_\nu m_\mu + (\alpha + \xbar\beta) m_\nu n_\mu - \lambda m_\nu m_\mu - \bar\mu m_\nu \bm_\mu.
\end{split}
\ee
For algebraically special case with $l$ affine parametrized, one obtains
\be\label{dN}
\begin{split}
N_{\nu\mu}\equiv & \n_\nu n_\mu= - (\gamma + \xbar\gamma) l_\nu n_\mu + \bar\pi n_\nu \bm_\mu + \pi n_\nu m_\mu  + \bar \nu l_\nu \bm_\mu + \nu l_\nu m_\mu  - \lambda m_\nu m_\mu \\
&  - \xbar\lambda \bm_\nu \bm_\mu  - \mu \bm_\nu m_\mu  - \bar\mu m_\nu \bm_\mu + (\alpha + \xbar\beta) m_\nu n_\mu + (\xbar\alpha + \beta) \bm_\nu n_\mu.
\end{split}
\ee
Then, the next group
\be
\begin{split}\label{m}
l^\nu \n_\nu m_\mu=&(\epsilon - \bar\epsilon) m_\mu - \kappa n_\mu + \bar\pi l_\mu,\\
n^\nu \n_\nu m_\mu= &(\gamma - \bar \gamma) m_\mu - \tau n_\mu + \xbar \nu l_\mu,\\
m^\nu \n_\nu m_\mu = &(\beta - \xbar\alpha) m_\mu - \sigma n_\mu + \xbar\lambda l_\mu,\\
\bm^\nu \n_\nu m_\mu= &(\alpha - \xbar \beta) m_\mu - \rho n_\mu + \xbar\mu l_\mu,\\
\n_\nu m_\mu =& (\epsilon - \bar\epsilon) n_\nu  m_\mu - \kappa  n_\nu  n_\mu + \bar\pi n_\nu  l_\mu  + (\gamma - \bar \gamma) l_\nu m_\mu - \tau l_\nu  n_\mu + \xbar \nu l_\nu  l_\mu \\
& - (\beta - \xbar\alpha) \bm_\nu m_\mu + \sigma \bm_\nu  n_\mu - \xbar\lambda  \bm_\nu l_\mu  - (\alpha - \xbar \beta) m_\nu m_\mu + \rho m_\nu n_\mu - \xbar\mu m_\nu l_\mu
\end{split}
\ee
For algebraically special case with $l$ affine parametrized, one obtains
\be\label{dM}
\begin{split}
M_{\nu\mu}\equiv  & \n_\nu m_\mu =(\gamma - \bar \gamma) l_\nu m_\mu + \bar\pi n_\nu  l_\mu - \tau l_\nu  n_\mu \\
&  + \xbar \nu l_\nu  l_\mu - \xbar\lambda  \bm_\nu l_\mu - (\beta - \xbar\alpha) \bm_\nu m_\mu \\
& - (\alpha - \xbar \beta) m_\nu m_\mu   + \rho m_\nu n_\mu - \xbar\mu m_\nu l_\mu,
\end{split}
\ee
and its complex conjugate
\be\label{dMb}
\begin{split}
\xbar M_{\nu\mu}\equiv & \n_\nu \bm_\mu = (\bar \gamma -  \gamma) l_\nu \bm_\mu + \pi n_\nu  l_\mu  - \xbar\tau l_\nu  n_\mu  +  \nu l_\nu  l_\mu - \lambda  m_\nu l_\mu  \\
& - ( \xbar\beta -\alpha) m_\nu \bm_\mu  - (\xbar \alpha - \beta) \bm_\nu \bm_\mu   + \xbar\rho \bm_\nu n_\mu - \mu \bm_\nu l_\mu.
\end{split}
\ee
As most of the Greeks are used for the spin coefficients, we will use Latins for the spacetime index to avoid notational confusion. The Weyl tensor can be constructed from the Weyl scalars as
%\begin{widetext}
\be\label{Weyl}
\begin{split}
& W_{abcd}=-\frac12(\Psi_2+\xbar\Psi_2)\big[\{l_a n_b l_c n_d\} + \{m_a \bm_b m_c \bm_d\}\big] + (\Psi_2 - \xbar\Psi_2) \{l_a n_b m_c \bm_d\}  \\
&+ \bigg[ - \frac12 \Psi_0 \{n_a \bm_b n_c \bm_d\}  - \frac12 \Psi_4 \{l_a m_b l_c m_d\} + \Psi_2 \{l_a m_b n_c \bm_d\}\\
& - \Psi_1 \left(\{l_a n_b n_c m_d\}+\{n_a \bm_b \bm_c m_d\}\right) + \Psi_3 \left(\{l_a n_b l_c m_d\} - \{l_a m_b m_c \bm_d\}\right) + c.c \bigg].
\end{split}
\ee
Note that the $c.c$ is only valid in the bracket $[]$ for the second and third lines and the notation $\{\}$ means the permutation of the index as the Weyl tensor
\be
\{abcd\}=abcd-bacd-abdc+badc + cdab - dcab -cdba + dcba.
\ee

For algebraically special solutions, $\Psi_0=0=\Psi_1$, the Weyl tensor is given by
\begin{align}
W_{abcd}=&-\frac12(\Psi_2+\xbar\Psi_2)\big[\{l_a n_b l_c n_d\} + \{m_a \bm_b m_c \bm_d\}\big] + (\Psi_2 - \xbar\Psi_2) \{l_a n_b m_c \bm_d\}  \label{W-type-II}\\
&\bigg[ - \frac12 \Psi_4 \{l_a m_b l_c m_d\} + \Psi_2 \{l_a m_b n_c \bm_d\} + \Psi_3 \left(\{l_a n_b l_c m_d\} - \{l_a m_b m_c \bm_d\}\right) + c.c \bigg].\nn
\end{align}
%\end{widetext}
For deriving the Weyl invariants, one needs to compute the contractions of the null bases. Then, the point is only about the orthogonality and normalization of the null bases. Hence, it is not necessary to know the real tetrads from the solution space. The final results must be the same for any group of four null vectors that satisfy \eqref{tetradcondition}. The information is only encoded in the Weyl scalars $\Psi$s and $\xbar\Psi$s. One can choose the simplest tetrads satisfying \eqref{tetradcondition} to compute the Weyl invariants in a much more efficient way. For algebraically special solutions, the square of the Weyl tensor is simply
\be
W_{abcd}W^{abcd}=24\left[(\Psi_2)^2 + (\xbar\Psi_2)^2\right].
\ee
The cubic of the Weyl tensor is 
\be
{W^{ab}}_{cd} {W^{cd}}_{ef} {W^{ef}}_{ab}=48\left[(\Psi_2)^3 + (\xbar\Psi_2)^3\right].
\ee
The 4th power of the Weyl tensor is 
\be
{W^{ab}}_{cd} {W^{cd}}_{ef} {W^{ef}}_{gh}{W^{gh}}_{ab}=288\left[(\Psi_2)^4 + (\xbar\Psi_2)^4\right].
\ee
It is also true that
\be
W^n \sim  \left[(\Psi_2)^n + (\xbar\Psi_2)^n\right],
\ee
The unknown prefactor can be fixed by, e.g., checking for the simplest Schwarzschild solution.

%%%%%%%%%%%%%%%%%%%%%%%%%%%%%%%%%%%%%%%%%%%%%%%%%%%%%%%%%%%%%%%%%%%%%%%%%%%%%%%%%%%%%%%%%%%%%%%%%%%%%%%%%%%%%%%%%%%%%%%%%%%%%%%%%%%%%%%%%%%%%%%%%%%%%%

\section{Square of the covariant derivative of the Weyl tensor}

Now we compute the invariants including one covariant derivative of the Weyl tensor. There are two types of contractions
\be
\n_e W_{abcd} \n^e W^{abcd}, \quad\quad \n_a W_{ebcd} \n^e W^{abcd}.
\ee
The strategy is as follow. The covariant derivative will act on both the Weyl scalars and the bases in \eqref{Weyl}. Thus the computation of the Weyl invariants must involve the solution space tetrad and $\n_a \Psi$. However, one can use the relation for the connection $L_{\nu\mu}$, $N_{\nu\mu}$, $M_{\nu\mu}$ $\xbar M_{\nu\mu}$ in \eqref{dL}, \eqref{dN}, \eqref{dM}, and \eqref{dMb} to replace the covariant derivative of the tetrad by the combination of the spin coefficients and the null bases. For the derivative on the Weyl scalar, we can use the directional derivative to replace the spacetime derivative to make everything as scalar times the tetrad bases. We have
\be
\n_\mu \Psi=D\Psi n_\mu + \Delta \Psi l_\mu - \delta \Psi \bm_\mu - \bar\delta \Psi m_\mu,
\ee
as well as the complex conjugate. With all the ingredients, one can repeat the strategy for Weyl invariants of polynomial, that is, choosing the simplest bases satisfying the orthogonality and normalization to compute the contraction. For example, the expression
\be
\begin{split}
&\n_e (\Psi_2 \{l_a m_b n_c \bm_d\})=\n_e \Psi_2  \{l_a m_b n_c \bm_d\} + \Psi_2 \n_e  (\{l_a m_b n_c \bm_d\})\\
&=(D\Psi_2 n_e + \Delta \Psi_2 l_e - \delta \Psi_2 \bm_e - \bar\delta \Psi_2 m_e) \{l_a m_b n_c \bm_d\}  + \Psi_2 \n_e  (\{l_a m_b n_c \bm_d\}),
\end{split}
\ee
can be treated as follows. For the second piece on the right hand, one should first write explicitly the permutation under $\{\}$, then act with $\n_e$ on every combination. In the end, there will be 32 terms. Finally, replace $\n l$ by $L$, $\n n$ by $N$, $\n m$ by $M$, $\n \bm$ by $\xbar M$. Then the whole expression is written as scalar times tetrad bases, there is no derivative. The contraction is only implemented among the bases which satisfy orthogonality and normalization conditions \eqref{tetradcondition}.

For algebraically general solutions, one can turn off $\Psi_0=0$ by tetrad rotation. Following the above algorithm, we obtain
\be
\begin{split}
&\n_e W_{abcd} \n^e W^{abcd}= 80 \left(D \Psi_2 \Delta \Psi_2 - \delta \Psi_2 \xbar\delta \Psi_2 \right)  + 160 \Psi_3 \left(\tau D \Psi_2 + \kappa \Delta \Psi_2 - \rho \delta \Psi_2 - \sigma \xbar\delta \Psi_2\right) \\
&  - 160 \Psi_1 \left(\nu D \Psi_2 + \pi \Delta \Psi_2 - \lambda \delta \Psi_2 - \mu \xbar\delta \Psi_2\right)   + 32 \Psi_1 \Big[\kappa \left(\Delta \Psi_4 + 4 \gamma \Psi_4 \right) - \sigma \left(\xbar\delta \Psi_4 + 4 \alpha \Psi_4\right) \Big] \\
&  - 160 \Psi_1 \Big[\rho \left(\Delta \Psi_3 + 2 \gamma \Psi_3 \right) - \tau \left(\xbar\delta \Psi_3 + 2 \alpha \Psi_3\right)\Big]  + 64\Psi_1 \Psi_3 \Big[ 5 (\mu  \rho - \pi \tau ) + 7(\lambda \sigma - \kappa \nu)\Big]\\
&  + 320 (\Psi_1)^2 (\nu \pi - \lambda \mu)+ 320 (\Psi_3)^2 (\kappa \tau - \rho \sigma) + 480 \Psi_1 \Psi_2 (\nu \rho -\tau \lambda)    + c.c. . 
\end{split}
\ee
We verify that the second invariant $\n_e W_{abcd} \n^a W^{ebcd}$ is a half of the above one. This is an intrinsic property of the Weyl tensor. The following relations from the Bianchi identity have been applied,
\be
\begin{split}
& D \Psi_1=2(2\rho + \epsilon) \Psi_1 - 3 \kappa \Psi_2,\\
&\xbar\delta \Psi_1 = D \Psi_2 - 2 (\pi -\alpha) \Psi_1  - 3\rho \Psi_2 + 2 \kappa \Psi_3,\\
& D \Psi_3 = \xbar\delta \Psi_2 - 2 \lambda \Psi_1 + 3 \pi \Psi_2 - 2(\epsilon - \rho) \Psi_3 - \kappa \Psi_4, \\
& D \Psi_4=\xbar \delta \Psi_3 - 3 \lambda \Psi_2 + 2 (2\pi+ \alpha ) \Psi_3- (4 \epsilon - \rho) \Psi_4 ,\\
& \delta \Psi_1 = 2 (2\tau + \beta) \Psi_1 - 3 \sigma \Psi_2,\\
&\Delta \Psi_1 = \delta \Psi_2 + 2 (\gamma - \mu)\Psi_1 - 3\tau \Psi_2 + 2 \sigma \Psi_3,\\
& \delta \Psi_3 = \Delta \Psi_2 - 2 \nu \Psi_1 +  3\mu \Psi_2 - 2 (\beta -\tau ) \Psi_3 - \sigma \Psi_4,\\
& \delta\Psi_4 = \Delta \Psi_3 - 3 \nu \Psi_2 + 2 (\gamma + 2\mu) \Psi_3 + (\tau -4\beta) \Psi_4.
\end{split}
\ee
A direct observation from the above results is that type-III and type-N solutions have vanishing Weyl invariants even for the case with one covariant derivative acting on the Weyl tensor. For algebraically special spacetime, we can further arrange the tetrads, such that $\Psi_1=0$, and $\kappa=0=\sigma$. Correspondingly, the Bianchi identity yields $\tau D \Psi_2 = \rho \delta \Psi_2$. Then the Weyl invariant is reduced to
\be
\begin{split}
\n_e W_{abcd} \n^e W^{abcd}
=& 80 \left(D \Psi_2 \Delta \Psi_2 + D \xbar\Psi_2 \Delta \xbar\Psi_2- \delta \Psi_2 \xbar\delta \Psi_2  - \xbar\delta \xbar\Psi_2 \delta \xbar\Psi_2\right)\\
=& 40\left(\n_\mu \Psi_2 \n^\mu \Psi_2 + \n_\mu \xbar\Psi_2 \n^\mu \xbar\Psi_2 \right). 
\end{split}
\ee
Alternatively,
\be
\n_e W_{abcd} \n^e W^{abcd}=
 240 \left(\rho \Psi_2 \Delta \Psi_2 + \xbar\rho \xbar\Psi_2 \Delta \xbar\Psi_2- \tau \Psi_2 \xbar\delta \Psi_2  - \xbar\tau \xbar\Psi_2 \delta \xbar\Psi_2\right). 
\ee
Hence, this Weyl invariant for the algebraic solution in the main text is
\be
\n_e W_{abcd} \n^e W^{abcd}=-360\left(\frac{L^2}{\varrho^8 }  + \frac{\bL^2}{\xbar\varrho^8} \right) \left(1 + \frac{L}{\varrho }  + \frac{\bL}{\xbar\varrho} \right).
\ee

\section{Invariants with one derivative in the solution space}

The simplest invariant with one derivative acting on $L$ and $\bL$ is
\begin{multline}
\n_\mu (W_2 )\n^\mu( W_2)=4608\bigg[ 3\left(\Psi_2 \Delta \Psi_2 + \xbar \Psi_2 \Delta \Psi_2\right)\left(\rho (\Psi_2)^2 + \xbar \rho (\xbar\Psi_2)^2 \right)  \\
- \left(\Psi_2 \xbar\delta \Psi_2 + 3 \xbar\tau (\xbar\Psi_2)^2\right)\left(\xbar\Psi_2 \delta \xbar\Psi_2 + 3 \tau (\Psi_2)^2\right)\bigg].
\end{multline}
Inserting the solution space yields
\begin{multline}
\n_\mu (W_2 )\n^\mu( W_2)=-20736\left(1+ \frac{L}{\varrho} + \frac{\bL}{\xbar\varrho} \right)\left(\frac{L^2}{\varrho^7} + \frac{\bL^2}{\xbar\varrho^7} \right)^2 \\
- 4608 \frac{L \bL}{\varrho^7 \xbar\varrho^7} \left(\eth \bL + \frac{6i \bL \eth \Sigma}{\xbar\varrho} \right) \left(\xbar\eth L - \frac{6i L \xbar\eth \Sigma}{\varrho} \right) .
\end{multline}
We use the ``$\eth$'' operator on the sphere \cite{Newman:1966ub} to simplify the formulas. 
For a field of spin $s$ as, they are defined as
\be
\begin{split}
&\eth \eta^{(s)}=(P\bp + 2 s \xbar\alpha^0)\eta^{(s)}=P\bP^{-s}\bp (\bP^s \eta^{(s)}) ,\\
&\xbar\eth \eta^{(s)}=(\bP\p - 2 s \alpha^0)\eta^{(s)}=\bP P^{s}\p (P^{-s} \eta^{(s)}) ,
\end{split}
\ee
where $(s)$ denotes the spin weight of the field $\eta^{(s)}$. For the case of unit sphere boundary, we have $P=\bP=P_S$. The operators $\eth$ and $\overline{\eth}$ will raise and lower the spin weight. The spin weights of relevant fields are listed in Table \ref{t1}. 
\begin{table}[ht]
\caption{Spin weights}\label{t1}
\begin{center}\begin{tabular}{|c|c|c|c|c|c|c|c|c|c|c|c|c|c|c|c|c|c|c|c|c}
\hline
& $\eth$ & $\tau_0$ & $\alpha^0$ & $\gamma^0$ & $\nu^0$ & $\mu^0$ & $\sigma^0$ & $\lambda^0$  & $\Psi^0_4$ &  $\Psi^0_3$ & $\Psi^0_2$ & $\Psi^0_1$ & $\Psi_0^0$  & $\phi_2^0$  & $\phi_1^0$ & $\phi_0^0$ \\
\hline
s & $1$& $1$& $-1$ &$0$& $-1$& $0$& $2$& $-2$  &
  $-2 $&  $-1$ & $0$ & $1$ & $2$ & $-1$ & $0$ & $1$ \\
\hline
\end{tabular}\end{center}\end{table}
In particular, 
\be
\xbar\eth \Sigma=P_S \p \Sigma, \quad\quad \xbar\eth^2 \Sigma=\p(P_S^2 \p\Sigma),\quad\quad \xbar\eth L=P_S \p L, \quad\quad \xbar\eth^2 L=\p(P_S^2 \p L).
\ee
A second class of invariant with one derivative acting on $L$ and $\bL$ is given by
\be
\begin{split}
&\n_\mu (dW_2 )\n^\mu(dW_2) \\
& = - 129600   \left(1+ \frac{L}{\varrho} + \frac{\bL}{\xbar\varrho} \right) \bigg[  8 \left(\frac{L^2}{\varrho^9} + \frac{\bL^2}{\xbar\varrho^9} \right) \left(1+\frac{L }{\varrho } + \frac{\bL }{\xbar\varrho } \right)  + \left(\frac{L^2}{\varrho^8} + \frac{\bL^2}{\xbar\varrho^8} \right)\left(\frac{L }{\varrho^2} + \frac{\bL }{\xbar\varrho^2} \right) \bigg]^2  \\
& -\frac{ 259200}{\varrho^2\xbar\varrho^2} \bigg[2 \left(1+ \frac{L}{\varrho} + \frac{\bL}{\xbar\varrho} \right)\left( \frac{\bL \eth \bL}{\xbar\varrho^7} + \frac{8i\bL^2 \eth \Sigma}{\xbar\varrho^8} \right)   + \left(\frac{L^2}{\varrho^8} + \frac{\bL^2}{\xbar\varrho^8} \right) \left(  \eth \bL + \frac{ 2i\bL \eth \Sigma}{\xbar\varrho} \right) \bigg]\\
&\quad\times \bigg[2 \left(1+ \frac{L}{\varrho} + \frac{\bL}{\xbar\varrho} \right)\left( \frac{ L \xbar\eth  L}{ \varrho^8} - \frac{8i L^2 \xbar\eth \Sigma }{ \varrho^7}\right)  + \left(\frac{L^2}{\varrho^8} + \frac{\bL^2}{\xbar\varrho^8} \right) \left(  \xbar\eth L -  \frac{ 2i L \xbar\eth \Sigma }{\varrho}\right) \bigg] .
\end{split}
\ee

\section{Invariants with two derivatives in the solution space}
\label{twoderivative}
The simplest invariant with two derivatives acting on $L$ and $\bL$ is
\be
\begin{split}
\n_\mu &  (\n_\nu W_2 \n^\nu W_2 ) \n^\mu(W_2)= \\
& 144 \left(1+ \frac{L}{\varrho} + \frac{\bL}{\xbar\varrho} \right) \left(\frac{L^2}{\varrho^7} + \frac{\bL^2}{\xbar\varrho^7}\right) \times \Bigg\{20736 \left(\frac{L}{\varrho^2} + \frac{\bL}{\xbar\varrho^2}\right) \left(\frac{L^2}{\varrho^7} + \frac{\bL^2}{\xbar\varrho^7}\right)^2  \\
&\qquad  + 290304 \left(1+ \frac{L}{\varrho} + \frac{\bL}{\xbar\varrho} \right) \times \left(\frac{L^2}{\varrho^7} + \frac{\bL^2}{\xbar\varrho^7}\right) \left(\frac{L^2}{\varrho^8} + \frac{\bL^2}{\xbar\varrho^8}\right) \\
& + 32256 \frac{L \bL}{\varrho^7 \xbar\varrho^7}\left(\frac{1}{\varrho} + \frac{1}{\xbar\varrho} \right) \left(\eth \bL + \frac{6i \bL \eth \Sigma}{\xbar\varrho} \right) \left(\xbar\eth L - \frac{6i L \xbar\eth \Sigma}{\varrho} \right)\\
&\qquad + 27648  \frac{L \bL}{\varrho^7 \xbar\varrho^7} \bigg[  \frac{i \bL \eth \Sigma}{\xbar\varrho^2} \left(\xbar\eth L - \frac{6i L \xbar\eth \Sigma}{\varrho} \right) - \frac{i L \xbar\eth \Sigma}{\varrho^2} \left(\eth \bL + \frac{6i \bL \eth \Sigma}{\xbar\varrho} \right)  \bigg]\Bigg\}\\
&+\frac{48}{\varrho\xbar\varrho} \Bigg\{ \frac{L}{\varrho^6}\left(\xbar\eth L - \frac{6i L \xbar\eth \Sigma}{\varrho} \right) \times \bigg[ 20736 \frac{1}{\xbar\varrho} \left(  \eth \bL + \frac{ 2i\bL \eth \Sigma}{\xbar\varrho} \right) \left(\frac{L^2}{\varrho^7} + \frac{\bL^2}{\xbar\varrho^7}\right)^2  \\
&\qquad  + 82944 \frac{\bL}{\xbar\varrho^7} \left(1+ \frac{L}{\varrho} + \frac{\bL}{\xbar\varrho} \right) \times \left(\frac{L^2}{\varrho^7} + \frac{\bL^2}{\xbar\varrho^7}\right)  \left(\eth \bL + \frac{7i \bL \eth \Sigma}{\xbar\varrho} \right) \\
& + 4608 \frac{L}{\varrho^7\xbar\varrho^7} \left(\eth \bL + \frac{14i \bL \eth \Sigma}{\xbar\varrho} \right) \left(\eth \bL + \frac{6i \bL \eth \Sigma}{\xbar\varrho} \right) \left(\xbar\eth L - \frac{6i L \xbar\eth \Sigma}{\varrho} \right) \\
&\qquad  + 4608 \frac{L\bL}{\varrho^7\xbar\varrho^7}  \left(\xbar\eth L - \frac{6i L \xbar\eth \Sigma}{\varrho} \right)  \left(\eth^2 \bL + \frac{6i \eth\bL \eth \Sigma}{\xbar\varrho} + \frac{6i \bL \eth^2 \Sigma}{\xbar\varrho} -  \frac{6  \bL (\eth \Sigma)^2}{\xbar\varrho^2}\right)\\
&\qquad+ 27648 \frac{L\bL}{\varrho^7\xbar\varrho^7} \left(\eth \bL + \frac{6i \bL \eth \Sigma}{\xbar\varrho} \right) \left( - \frac{i L \eth\xbar\eth \Sigma}{\varrho} + \frac{  L \eth\Sigma \xbar\eth \Sigma}{\varrho^2}\right) \bigg]+ c.c \Bigg\}.
\end{split}
\ee
A second class of invariant with two derivatives acting on $L$ and $\bL$ is given by
\begin{align}
&\n_\mu \n_\nu (W_2 )\n^\mu \n^\nu ( W_2)= \nn \\
& 2 \Big[ \xbar\delta \xbar\delta W_2 + (\alpha-\xbar\beta)\xbar\delta W_2 - \lambda D W_2 + \xbar\sigma \Delta W_2\Big] \times  \Big[ \delta \delta W_2 + (\xbar\alpha-\beta) \delta W_2  - \xbar\lambda D W_2 + \sigma \Delta W_2 \Big] \nn\\
& + 2 \Big[\xbar\delta \delta W_2 - (\alpha - \xbar\beta) \delta W_2 - \xbar\mu D W_2 + \rho \Delta W_2 \Big] \times 
\Big[\delta \xbar\delta W_2 - (\xbar\alpha - \beta) \xbar\delta W_2 - \mu D W_2 + \xbar\rho \Delta W_2 \Big] \nn\\
& + 2 \Big[ D D W_2 - (\epsilon + \xbar \epsilon) D W_2 + \kappa \xbar\delta W_2 + \xbar\kappa  \delta W_2 \Big]   \times \Big[ \Delta \Delta W_2 + (\gamma + \xbar \gamma) \Delta W_2 - \nu \delta W_2 - \xbar \nu \xbar\delta W_2\Big] \nn\\
& + 2 \Big[D\Delta W_2 +  (\epsilon + \xbar \epsilon) \Delta W_2 - \pi \delta W_2 - \xbar\pi \xbar\delta W_2 \Big]\times \Big[\Delta D W_2 -  (\gamma + \xbar \gamma) D W_2 + \tau \xbar\delta W_2 + \xbar \tau \delta W_2 \Big] \nn\\
& -2 \Big[\delta \Delta W_2 + (\xbar\alpha + \beta) \Delta W_2 - \xbar\lambda \xbar\delta W_2 - \mu \delta W_2 \Big] \times \Big[\xbar\delta D W_2 - (\alpha + \xbar \beta) D W_2 + \rho \xbar\delta W_2 + \xbar\sigma \delta W_2\Big] \\
& -2 \Big[\xbar\delta \Delta W_2 + (\alpha + \xbar\beta) \Delta W_2 - \lambda \delta W_2 - \xbar\mu \xbar\delta W_2 \Big] \times \Big[\delta D W_2 - (\xbar\alpha + \beta) D W_2 +\xbar \rho \delta W_2 + \sigma \xbar\delta W_2\Big] \nn\\
& - 2 \Big[D \delta W_2 - (\epsilon - \xbar \epsilon) \delta W_2 - \xbar\pi D W_2 + \kappa \Delta W_2 \Big]\times \Big[\Delta \xbar \delta W_2 + (\gamma - \xbar \gamma) \xbar \delta W_2 - \nu D W_2 + \xbar \tau \Delta W_2 \Big] \nn\\
& -2 \Big[D \xbar \delta W_2 + (\epsilon - \xbar \epsilon) \xbar\delta W_2 - \pi D W_2  + \xbar\kappa \Delta W_2\Big]\times \Big[ \Delta \delta W_2 - (\gamma - \xbar \gamma) \delta W_2 - \xbar \nu D W_2 + \tau \Delta W_2\Big].\nn
\end{align}
Inserting the solutions yields
\begin{align}
&\n_a \n_b (W_2 )\n^a \n^b ( W_2)= \frac12 \p_r \left[\left(1+ \frac{L}{\varrho} + \frac{\bL}{\xbar\varrho} \right) \p_r W_2 \right]  \left(1+ \frac{L}{\varrho} + \frac{\bL}{\xbar\varrho} \right) \p_r \p_r W_2 \nn\\
&\qquad - \frac12 \p_r \left[\left(1+ \frac{L}{\varrho} + \frac{\bL}{\xbar\varrho} \right) \p_r W_2 \right] 
\left(\frac{L}{\varrho^2} + \frac{\bL}{\xbar\varrho^2}\right) \p_r W_2 + \frac{2}{\varrho\xbar\varrho} \bigg[ \xbar\eth\big[\frac{1}{\varrho}(\xbar\eth W_2 + i \xbar\eth \Sigma \p_r W_2)\big] \nn\\
& + i \xbar\eth\Sigma \p_r \big[\frac{1}{\varrho}(\xbar\eth W_2 + i \xbar\eth \Sigma \p_r W_2)\big] \bigg] \bigg[ \eth\big[\frac{1}{\xbar\varrho}( \eth W_2 - i  \eth \Sigma \p_r W_2)\big] - i \eth\Sigma \p_r \big[\frac{1}{\xbar\varrho}( \eth W_2 - i \eth \Sigma \p_r W_2)\big] \bigg]\nn\\
&\qquad + \p_r \big[\frac{1}{\xbar\varrho}( \eth W_2 - i \eth \Sigma \p_r W_2)\big] \bigg[ \left(1+ \frac{L}{\varrho} + \frac{\bL}{\xbar\varrho} \right) \p_r \big[\frac{1}{\varrho}(\xbar \eth W_2 + i \xbar \eth \Sigma \p_r W_2)\big]\nn\\
&\qquad \qquad +\left( \frac{\xbar\eth L}{\varrho^2} - \frac{2 i L \xbar\eth \Sigma}{\varrho^3}\right) \p_r W_2 + \left(\frac{L}{\varrho^2} - \frac{\bL}{\xbar\varrho^2} \right) \big[\frac{1}{\varrho}(\xbar\eth W_2 + i \xbar\eth \Sigma \p_r W_2)\big]\bigg]\nn\\
&\qquad + \p_r \big[\frac{1}{\varrho}( \xbar\eth W_2 + i \xbar\eth \Sigma \p_r W_2)\big] \bigg[ \left(1+ \frac{L}{\varrho} + \frac{\bL}{\xbar\varrho} \right) \p_r \big[\frac{1}{\xbar\varrho}( \eth W_2 - i  \eth \Sigma \p_r W_2)\big]\nn\\
&\qquad \qquad +\left( \frac{\eth \bL}{\xbar\varrho^2} + \frac{2 i \bL \eth \Sigma}{\xbar\varrho^3}\right) \p_r W_2 - \left(\frac{L}{\varrho^2} - \frac{\bL}{\xbar\varrho^2} \right) \big[\frac{1}{\xbar\varrho}(\eth W_2 - i \eth \Sigma \p_r W_2)\big]\bigg]\nn\\
&\qquad + \p_r \p_r W_2 \bigg[ \frac12 \left(\frac{L}{\varrho^2} + \frac{\bL}{\xbar\varrho^2} \right) \left(1+ \frac{L}{\varrho} + \frac{\bL}{\xbar\varrho} \right) \p_r W_2 + \frac12 \left(1+ \frac{L}{\varrho} + \frac{\bL}{\xbar\varrho} \right) \nn\\
&\qquad \times \p_r \left(\left(1+ \frac{L}{\varrho} + \frac{\bL}{\xbar\varrho} \right) \p_r W_2 \right) -\left( \frac{\xbar\eth L}{\varrho^2} - \frac{2 i L \xbar\eth \Sigma}{\varrho^3}\right) \big[\frac{1}{\xbar\varrho}(\eth W_2 - i \eth \Sigma \p_r W_2)\big]  \nn\\
&\qquad  - \left( \frac{\eth \bL}{\xbar\varrho^2} + \frac{2 i \bL \eth \Sigma}{\xbar\varrho^3}\right) \big[\frac{1}{\varrho}(\xbar\eth W_2 + i \xbar\eth \Sigma \p_r W_2)\big] \bigg]\nn\\
&\qquad + \bigg(\frac{1}{\xbar\varrho}(\eth  - i \eth \Sigma \p_r ) \big[\left(1+ \frac{L}{\varrho} + \frac{\bL}{\xbar\varrho} \right) \p_r W_2 \big] -\left(\frac{1}{\xbar\varrho} + \frac{2rL}{\xbar\varrho \varrho^2}\right) \big[\frac{1}{\xbar\varrho}(\eth W_2 - i \eth \Sigma \p_r W_2)\big] \bigg)\nn\\
&\qquad \qquad \times \left( \frac{1}{\varrho}(\xbar\eth \p_r W_2 + i \xbar\eth \Sigma \p_r \p_r W_2) - \frac{1}{\varrho^2} (\xbar\eth W_2 + i \xbar\eth \Sigma \p_r W_2)   \right) \nn\\ 
&\qquad + \bigg(\frac{1}{\varrho}(\xbar\eth  + i \xbar\eth \Sigma \p_r ) \big[\left(1+ \frac{L}{\varrho} + \frac{\bL}{\xbar\varrho} \right) \p_r W_2 \big] -\left(\frac{1}{\varrho} + \frac{2r\bL}{\varrho \xbar\varrho^2}\right) \big[\frac{1}{\varrho}(\xbar\eth W_2 + i \xbar\eth \Sigma \p_r W_2)\big] \bigg)\nn\\
&\qquad \qquad \times \left( \frac{1}{\xbar\varrho}(\eth \p_r W_2 - i \eth \Sigma \p_r \p_r W_2) - \frac{1}{\xbar\varrho^2} (\eth W_2 - i \eth \Sigma \p_r W_2)   \right) \nn\\
&\qquad + 2 \bigg[ \frac{1}{\varrho}(\xbar\eth  + i \xbar\eth \Sigma \p_r ) \big[\frac{1}{\xbar\varrho}(\eth W_2 - i \eth \Sigma \p_r W_2)\big] + \left(\frac{1}{2\varrho} + \frac{r\bL}{\varrho \xbar\varrho^2}\right) \p_r W_2 \nn\\ 
&\qquad +  \frac{1}{2\varrho} \left(1+ \frac{L}{\varrho} + \frac{\bL}{\xbar\varrho} \right) \p_r W_2 \bigg] \times \bigg[ \frac{1}{\xbar\varrho}(\eth -  i \eth \Sigma \p_r ) \big[\frac{1}{\varrho}(\xbar\eth W_2 + i \xbar\eth \Sigma \p_r W_2)\big] \nn\\
&\qquad  + \left(\frac{1}{2\xbar\varrho} + \frac{r L}{\xbar\varrho \varrho^2}\right) \p_r W_2 +  \frac{1}{2\xbar\varrho} \left(1+ \frac{L}{\varrho} + \frac{\bL}{\xbar\varrho} \right) \p_r W_2 \bigg].
\end{align}
%\end{widetext}

\section{NP formalism for algebraically special spacetime including Maxwell fields}
\label{NPMaxwell}

In the NP formalism, the Maxwell-tensor is replaced by three complex scalars
\be
\begin{split}
\phi_0=F_{\mu\nu} l^\mu m^\nu,\quad \phi_1=\frac12 F_{\mu\nu} (l^\mu n^\nu + \bm^\mu m^\nu),\quad \phi_2=F_{\mu\nu} \bm^\mu n^\nu .
\end{split}
\ee
Conversely, the Maxwell tensor is
\be\label{Maxwell}
F_{\mu\nu}=2(\phi_1+\xbar\phi_1)\, n_{[\mu} l_{\nu]} + 2(\phi_1 - \xbar\phi_1) \, m_{[\mu} \bm_{\nu]} 
+ 2\phi_2\, l_{[\mu} m_{\nu]} + 2\xbar\phi_2\, l_{[\mu} \bm_{\nu]} .
\ee
The Lagrangian of four-dimensional Einstein-Maxwell theory in the usual NP formalism is \cite{Chandrasekhar}
\be
\cL=\sqrt{-g}\left[ R - \frac12 F^2\right],\qquad F=\td A.
\ee 

When we consider algebraically special solution, we consider the principle null direction of the Maxwell field align with the Weyl tensor, namely $\phi_0=0$. The tetrad system can be again constructed as
\be\label{gaugetetrad}
\begin{split}
&n=W\frac{\p}{\p u} + U \frac{\p}{\p r} + X^A \frac{\p}{\p x^A},\qquad
l=\frac{\p}{\p r},\\
&m= M\frac{\p}{\p u} + \omega\frac{\p}{\p r} + L^A \frac{\p}{\p x^A}.
\end{split}
\ee
The fall-off conditions are chosen as follows
\be\label{boundaryconditions}
\begin{split}
&W=1+\cO(r^{-1}),\quad X^A=\cO(r^{-1}),\quad U=\cO(r), \quad M=\cO(r^{-1}),\\  &L^z=\cO(r^{-2}),\quad L^{\bz}=\cO(r^{-1}),\quad \omega=\cO(r^{-1}), \\
& \rho=-\frac1r+\cO(r^{-2}),\quad Re(\rho)=-\frac{1}{r}+\cO(r^{-3}),\\
& \sigma=\cO(r^{-2}), \quad \alpha=\cO(r^{-1}), \quad \beta=\cO(r^{-1})\\
&\xbar\alpha+\beta=\cO(r^{-2}),\quad \lambda=\cO(r^{-1}),\quad \mu=\cO(r^{-1}), \quad \tau=\cO(r^{-1}).
\end{split}
\ee
Those conditions can be reached for solution space by three classes of tetrad rotations and the combined coordinates transformation which preserves the gauge conditions $\kappa=\epsilon=\pi=0$, see, e.g., the arrangement for some of fall-off conditions in \cite{Newman:1962cia}. The full NP equations including Maxwell fields with respect to conditions $\sigma=\kappa=\epsilon=\pi=\Psi_0=\Psi_1=\phi_0=0$ are arranged as

\textbf{Radial equations}
\bea
&&D\rho =\rho^2,\label{gR1}\\
&&D\alpha=\rho  \alpha ,\label{gR3}\\
&&D\beta  = \xbar\rho  \beta,\label{gR4}\\
&&D\tau =\tau \rho,\label{gR5}\\
&&D\lambda=\rho\lambda,\label{gR6}\\
&&D\mu =\xbar\rho \mu + \Psi_{2},\label{gR7}\\
&&D\gamma=\tau \alpha +  \xbar \tau \beta  + \Psi_2 + \phi_1 \xbar\phi_1,\label{gR8}\\
&&D\nu =\xbar\tau \mu + \tau  \lambda + \Psi_3 + \phi_2 \xbar\phi_1,\label{gR9}\\
&&D\Delta= - (\gamma + \xbar\gamma) D + \xbar\tau \delta + \tau \xbar\delta,\\
&&D\delta= - (\xbar\alpha +\beta) D + \xbar\rho \delta ,\\
&&D\Psi_2  =   3\rho \Psi_2 + 2\rho \phi_1 \xbar\phi_1,\label{gR16}\\
&&D\Psi_3 - \xbar\delta \Psi_2 =  2\rho \Psi_3 +\xbar \phi_1 D \phi_2,\label{gR17}\\
&&D\Psi_4 - \xbar\delta \Psi_3 = \rho  \Psi_4 + 2 \alpha \Psi_3 - 3 \lambda \Psi_2 + \bar\delta (\phi_2\xbar\phi_1)  + 2(\alpha-\xbar\tau)\phi_2\xbar\phi_1 - 2 \lambda\phi_1\xbar\phi_1,\label{gR18}\\
&&D \phi_1 - \bar\delta\phi_0=-2\alpha \phi_0 + 2 \rho \phi_1,\label{gR19}\\
&&D \phi_2 - \bar\delta \phi_1 = - \lambda \phi_0+\rho\phi_2\label{gR20}.
\eea
\textbf{Non-radial  equations}
\bea
&&\Delta\lambda  = \xbar\delta\nu- (\mu + \xbar\mu)\lambda - (3\gamma - \xbar\gamma)\lambda + \nu ( 3\alpha + \xbar\beta -\xbar\tau ) - \Psi_4,\label{gH1}\\
&&\Delta\rho= \xbar\delta\tau- \rho\xbar\mu   - \tau (\xbar\tau + \alpha -\xbar\beta) + (\gamma + \xbar\gamma)\rho  - \Psi_2 ,\label{gH2}\\
&&\Delta\alpha = \xbar\delta\gamma +\rho \nu - (\tau + \beta)\lambda + (\xbar\gamma -\xbar \mu)\alpha  + \gamma(\xbar\beta - \xbar \tau)  -\Psi_3 ,\label{gH3}\\
&&\Delta \mu=\delta\nu-\mu^2 - \lambda\xbar\lambda - (\gamma + \xbar\gamma)\mu   +  \nu (\xbar\alpha + 3\beta - \tau) - \phi_2 \xbar\phi_2,\label{gH4}\\
&&\Delta \beta=\delta\gamma - \mu\tau  + \beta(\gamma - \xbar\gamma -\mu) - \alpha\xbar\lambda  + \gamma (\xbar\alpha + \beta -\tau) - \phi_1 \xbar\phi_2,\label{gH5}\\
&&0=\delta\tau - \rho\xbar\lambda - \tau (\tau -\xbar\alpha + \beta) ,\label{gH6}\\
&&\Delta \delta=\delta \Delta + \xbar\nu D + (\xbar\alpha + \beta -\tau) \Delta  + (\gamma-\xbar\gamma -\mu)\delta - \xbar\lambda \xbar\delta,\label{gH7}\\
&&\delta\rho=\rho(\xbar\alpha + \beta)  + \tau (\rho - \xbar\rho) ,\label{gH9}\\
&&\delta\alpha - \xbar\delta\beta=\mu\rho  + \alpha\xbar\alpha + \beta\xbar\beta - 2 \alpha\beta + \gamma(\rho - \xbar\rho) - \Psi_2 + \phi_1 \xbar\phi_1 ,\label{gH10}\\
&&\delta\lambda - \xbar\delta\mu= \nu(\rho - \xbar\rho) + \mu (\alpha + \xbar\beta) + \lambda (\xbar\alpha - 3\beta) - \Psi_3 + \phi_2 \xbar\phi_1 ,\label{gH11}\\
&&\xbar\delta \delta - \delta \xbar\delta=(\xbar\mu-\mu) D + (\xbar\rho-\rho)\Delta + (\alpha -\xbar\beta) \delta + (\beta - \xbar\alpha)\xbar\delta,\label{gH12}\\ 
&& \delta \Psi_2 = 3\tau \Psi_2  + 2 \rho \phi_1 \xbar\phi_2 - 2 \tau \phi_1 \xbar\phi_1 ,\label{gH15}\\
&&\Delta\Psi_2 - \delta \Psi_3 =  - 3\mu \Psi_2 + (2\beta - 2\tau) \Psi_3   - D(\phi_2\xbar\phi_2) \nn\\ 
&& \qquad \qquad + \delta(\phi_2\xbar\phi_1) + 2 \beta \phi_2\xbar\phi_1 - 2\mu\phi_1\xbar\phi_1 + \xbar\rho \phi_2\xbar\phi_2,\label{gH16}\\
&&\Delta\Psi_3 - \delta \Psi_4 = 3\nu \Psi_2 - (2\gamma + 4\mu) \Psi_3 + (4\beta - \tau) \Psi_4 +\Delta(\phi_2\xbar\phi_1) - \bar\delta(\phi_2\xbar\phi_2)  \nn \\
&& \qquad \qquad  + 2(\xbar\mu+\gamma)\phi_2\xbar\phi_1 - 2\nu \phi_1\xbar\phi_1  + 2\lambda \phi_1\xbar\phi_2 +(\xbar\tau-2\alpha-2\xbar\beta)\phi_2\xbar\phi_2,\label{gH17}\\
&& \delta \phi_1= 2\tau \phi_1 ,\label{gH18}\\
&&\Delta \phi_1 - \delta \phi_2 =  - 2\mu \phi_1 - (\tau-2\beta)\phi_2. \label{gH19}
\eea

\section{Stationary algebraically special solutions}
\label{ASS}

In this section, we will derive the generic solution space of Einstein-Maxwell theory with respect to stationary algebraically special conditions and locally asymptotic flatness. The solutions to the radial equations are given by
\begin{align}
    &\rho=-\frac{1}{r+i\Sigma},\quad \alpha=\frac{\alpha^0}{r+i\Sigma},\quad \beta=-\xbar\alpha,\quad \tau=\frac{\tau^0}{r+i\Sigma},\quad \lambda=\frac{\lambda^0}{r+i\Sigma},\nn \\
    & \phi_1=\frac{\phi_1^0}{(r+i\Sigma)^2}, \quad \phi_2=\frac{\phi_2^0}{r+i\Sigma} - \frac{\xbar\eth_n \phi_1^0}{(r+i\Sigma)^2} + \frac{\phi_1^0 ( i \xbar\eth_n \Sigma + \xbar\omega^0)}{(r+i\Sigma)^3}, \nn\\
    &M=\frac{M^0}{r-i\Sigma},\quad \omega=\frac{\omega^0}{r-i\Sigma},\quad L^{\bz}=\frac{P}{r-i\Sigma},\nn\\
    &W=1-\frac{\xbar\tau^0 M^0}{r-i\Sigma}-\frac{\tau^0\xbar M^0}{r+i\Sigma},\quad X^z=-\frac{\tau^0\bP}{r+i\Sigma},\label{generaltypetwosolutionr}\\
    &\Psi_2=\frac{\Psi_2^0}{(r+i\Sigma)^3} + \frac{2\phi_1^0 \xbar \phi_1^0}{(r-i\Sigma)(r+i\Sigma)^3},\quad \mu=\frac{\mu^0}{r-i\Sigma}-\frac{r\Psi_2^0 + \phi_1^0\xbar\phi_1^0}{(r+i\Sigma)^2(r-i\Sigma)},\nn\\ 
    &\gamma=\gamma^0 + \frac{\xbar\tau^0\xbar\alpha^0}{r-i\Sigma} - \frac{\tau^0\alpha^0}{r+i\Sigma}-\frac{\Psi_2^0}{2(r+i\Sigma)^2} -\frac{\phi_1^0\xbar\phi_1^0}{(r+i\Sigma)^2(r-i\Sigma)},\nn\\ 
    &U=-(\gamma^0+\xbar\gamma^0)r+U^0-\frac{\Psi_2^0}{2(r+i\Sigma)}-\frac{\xbar\Psi_2^0}{2(r-i\Sigma)} - \frac{\tau^0\xbar\omega^0}{r+i\Sigma}-\frac{\xbar\tau^0\omega^0}{r-i\Sigma} - \frac{\phi_1^0 \xbar\phi_1^0}{r^2+\Sigma^2} .\nn 
\end{align}
Correspondingly, the co-tetrad system is given by
\begin{align}
&l=du - \frac{\xbar M^0}{\bP} dz - \frac{M^0}{P} d\bz,\nn\\
&n=\left[(\gamma^0+\xbar\gamma^0) r - U^0 + \frac{rK+J\Sigma+\phi_1^0\xbar\phi_1^0}{r^2+\Sigma^2} \right] l + dr  - \frac{\xbar\omega^0 }{\bP} d z  -  \frac{\omega^0 }{P} d \bz,\\
&m=-\tau^0 du + \left(-\frac{r}{\bP} + \frac{ \xbar M^0 \tau^0 - i\Sigma}{\bP}\right) dz + \frac{M^0 \tau^0}{P} d\bz,\nn\\
&\bm=-\xbar\tau^0 du + \frac{\xbar M^0 \xbar\tau^0 }{\bP} d z + \left(-\frac{r}{P} + \frac{M^0 \xbar\tau^0 + i\Sigma }{ P}\right) d\bz ,\nn
\end{align}
where $K$ and $J$ are the real and imaginary parts of $\Psi_2^0$,
\be
K=\frac12(\Psi_2^0 + \xbar\Psi_2^0),\quad\quad J=-\frac{i}{2}(\Psi_2^0 - \xbar\Psi_2^0).\nn
\ee
There are three more radial equations \eqref{gR9}, \eqref{gR17}, and \eqref{gR18} which can be in principle solved in the same way. We found that the solutions to those three equations will be much simplified when some of the non-radial equations are solved, in particular for the stationary case. The reason is that those three equations are relevant to the Bianchi identity. Only when the Maxwell's equations are satisfied, the stress tensor of the Maxwell fields is conserved, which suggests that one should solve all components of the Maxwell's equations before the Bianchi identities.

For the non-radial equations, all the information is expected to be encoded in the leading pieces of the $\frac{1}{r}$ expansion \cite{Newman:1962cia}, which was precisely verified for the algebraically special case for pure gravity case \cite{Mao:2024jpt}. We expect this should also hold for Einstein-Maxwell theory. We will first check the constraints from the leading orders of all the non-radial equations and then verify that there is no more constraint from any sub-leading order. Starting from the $\bz$-component and $r$-component of \eqref{gH7} lead to
\be
\gamma^0=-\frac12 \p_u \ln \bP=0,
\ee
for the stationary case and
\be
\nu^0=\xbar\eth (\gamma^0+\xbar\gamma^0)-\xbar\tau^0(\gamma^0+\xbar\gamma^0)=0.
\ee
We continue with the $u$-component of \eqref{gH7}, which yields
\be
\tau^0=-\p_u M^0 - 2 M^0 \xbar\gamma^0=0. 
\ee
From \eqref{gH9}, we obtain that
\be
\omega^0=2i\Sigma \tau^0- i\eth_n \Sigma=- i\eth \Sigma.
\ee
Then, \eqref{gH6} yields that
\be\label{lambda0}
\lambda^0=(\xbar\tau^0)^2-\xbar\eth_n \xbar\tau^0=0.
\ee
From the $u$-component of \eqref{gH12}, one can derive that
\be
\begin{split}
2i\Sigma=\eth \xbar M^0 - \xbar \eth M^0 
=-\frac{\left[P(\bP W - \xbar M^0 X^z)-M^0 \bP X^{\bz}\right]^2}{P\bP} l_{[u}\p_z l_{\bz]}.
\end{split}
\ee
Now, it is clear that $l$ is hypersurface orthogonal when $\Sigma=0$. From $\bz$-component of \eqref{gH12}, it is obtained that
\be
\alpha^0=\frac12 (\xbar M^0 \p_u \ln P + \bP \p \ln P)=\frac12 \bP \p \ln P.
\ee
From \eqref{gH10}, we get
\be
\mu^0=2i\Sigma \gamma^0+\mu^0_r=\mu^0_r,
\ee
where $\mu_r^0=-\eth\alpha^0-\xbar\eth \xbar\alpha^0=-\frac12 P\bP \p\bp \ln (P\bP)$ is the real part. Then one can verify the commutator of the operator $\eth$
\be\label{commutator}
[\xbar\eth,\eth]\eta^{(s)} =-2s\mu_r^0 \eta^{(s)}.
\ee
From \eqref{gH2}, we obtain
\be
U^0=\xbar\eth_n\tau^0-i\Sigma(\gamma^0 + 3 \xbar\gamma^0) - i\p_u \Sigma + \mu_r^0 - \tau^0\xbar\tau^0=\mu_r^0.
\ee
The $r$-component of \eqref{gH12} fixed the imaginary part of $\Psi_2^0$ as
\be\label{c}
\Psi_2^0-\xbar\Psi_2^0=i\Sigma (2U^0+\mu^0 + \xbar\mu^0) + \xbar\eth \omega^0 - \eth \xbar\omega^0.
\ee
From the Maxwell's equations \eqref{gH18} and \eqref{gH19}, we obtain
\be
\eth \phi_1^0 = 2 \tau^0 \phi_1^0=0,
\ee
and 
\be
\p_u \phi_1^0 + 2(\gamma^0 + \xbar\gamma^0)\phi_1^0=\eth \phi_2^0 - \tau^0 \phi_2^0,
\ee
which is reduced to $\eth \phi_2^0=0$ for the stationary case. Then we can solve the radial equation \eqref{gR9}, \eqref{gR17}, and \eqref{gR18}. The solutions are
\be
\begin{split}
&\nu = \frac{\xbar\eth \Psi_2^0}{2(r+i\Sigma)^2} - \frac{ i \Psi_2^0  \xbar\eth \Sigma}{(r+i\Sigma)^3} + \frac{\xbar\phi_1^0 \xbar\eth\phi_1^0}{(r+i\Sigma)^2(r-i\Sigma)}  - \frac{2i \phi_1^0 \xbar\phi_1^0 \xbar\eth \Sigma}{(r+i\Sigma)^3(r-i\Sigma)},\\
&\Psi_3=-\frac{\xbar\eth \Psi_2^0}{(r+i\Sigma)^3} + \frac{3 i\Psi_2^0 \xbar\eth \Sigma}{(r+i\Sigma)^4} - \frac{2\xbar\phi_1^0 \xbar\eth\phi_1^0}{(r+i\Sigma)^3(r-i\Sigma)} + \frac{6i \phi_1^0 \xbar\phi_1^0 \xbar\eth \Sigma}{(r+i\Sigma)^4(r-i\Sigma)},\\
&\Psi_4= \frac{\xbar\eth^2 \Psi_2^0}{2(r+i\Sigma)^3} - \frac{3i\xbar\eth\Psi_2^0 \xbar\eth \Sigma + i \Psi_2^0 \xbar\eth^2 \Sigma }{(r+i\Sigma)^4} - \frac{6\Psi_2^0  (\xbar\eth \Sigma)^2}{(r+i\Sigma)^5}  + \frac{\xbar\phi_1 \xbar\eth^2 \phi_1^0}{(r+i\Sigma)^3(r-i\Sigma)}\\
&\qquad- \frac{6i\xbar\phi_1^0 \xbar\eth\phi_1^0\xbar\eth \Sigma + 2i\xbar\phi_1^0  \phi_1^0 \xbar\eth^2 \Sigma }{(r+i\Sigma)^4(r-i\Sigma)} - \frac{12\xbar\phi_1^0 \phi_1^0 (\xbar\eth \Sigma)^2}{(r+i\Sigma)^5(r-i\Sigma)}.   
\end{split}
\ee
Eq. \eqref{gH11} yields that
\be
\Psi_3^0=2i\Sigma \nu^0 + \xbar\eth \mu^0 - \eth_n \lambda^0=\xbar\eth \mu^0.
\ee
From \eqref{gH1}, we obtain
\be
\Psi_4^0=\xbar\eth_n \nu^0 - 4\gamma^0\lambda^0 - \xbar\tau^0\nu^0 - \p_u\lambda^0=0.
\ee
A constraint of $\Psi_2^0$ is obtained from \eqref{gH15} as
\be\label{p}
\eth \Psi_2^0=3\tau^0\Psi_2^0 - 2 \phi_1^0 \xbar\phi_2^0=- 2 \phi_1^0 \xbar\phi_2^0.
\ee
From \eqref{gH16}, there is another constraint of $\Psi_2^0$,
\be\label{u}
\p_u \Psi_2^0 + 3 (\gamma^0 + \xbar\gamma^0) \Psi_2^0=\eth \Psi_3^0 - 2\tau^0 \Psi_3^0 + \phi_2^0 \xbar \phi_2^0,
\ee
which is reduced to $\eth\xbar\eth \mu^0=-\phi_2^0 \xbar \phi_2^0$ for stationary case.

\subsection{Reductions with a sphere boundary}

\label{general}

Now we implement the constraint that the boundary is a punctured unit sphere. The boundary line element is $\td s^2=\frac{4}{(1+z\bz)^2}\td z \td\bz$. Correspondingly, we have $\mu^0=-\frac12$ which fixes $\phi_2^0=0$ from \eqref{u}. The whole system is further simplified and described by a co-tetrad system 
\begin{align}
&l=\td u - \frac{\xbar \cM}{P_S} \td z - \frac{\cM}{P_S} \td \bz,\nn\\
&n= \td r  - \frac{\xbar\omega^0 }{P_S} \td z  -  \frac{\omega^0 }{P_S} \td \bz + \left[ \frac12 + \frac{rK+J\Sigma + \phi_1^0 \xbar\phi_1^0}{r^2+\Sigma^2} \right] l ,\\
&m=-\frac{1}{P_S} \left(r + i\Sigma \right) \td z ,\quad\quad \bm= -\frac{1}{P_S} \left(r - i\Sigma \right) \td \bz ,\nn
\end{align}
where we rename $M^0$ as $\cM$ which is an arbitrary complex function of variables $(z,\bz)$. We define that $K$ and $J$ are the real and imaginary parts of the arbitrary complex function $\Psi_2^0$ of variables $(z,\bz)$,
\be
\begin{split}
& K=\frac12\left[\Psi_2^0(z,\bz) + \xbar\Psi_2^0(z,\bz)\right],\\
& J=-\frac{i}{2}\left[\Psi_2^0(z,\bz) - \xbar\Psi_2^0(z,\bz)\right].
\end{split}
\ee
$\Sigma$ is the twist function which is fixed by $\cM$ from
\be
\Sigma=-\frac{i P_S^2}{2}\left[ \bp (\frac{\xbar \cM}{P_S} )- \p (\frac{  \cM}{P_S}) \right]. 
\ee
Moreover,
\be
\omega^0=- i P_S \bp \Sigma, \quad\quad \xbar\omega^0= i P_S \p \Sigma.
\ee
The unknown complex functions $\cM$, $\Psi_2^0$, $\phi_1^0$ and their complex conjugate are constrained by
\be
\begin{split}
&\Psi_2^0-\xbar\Psi_2^0=  -2 i (\Sigma + P^2_S \p\bp \Sigma), \\
& \bp \Psi_2^0=0, \quad \p \xbar\Psi_2^0=0 ,\quad \bp \phi_1^0=0,\quad \p \xbar\phi_1^0=0.
\end{split}
\ee
These equations comprise all the constraints from the Einstein equation, Maxwell's equations and Bianchi identities. The above construction covers all the algebraically-special stationary solutions with a boundary topology $\mathbb{R}\times \mathbb{C}_*$ under locally asymptotically flat conditions in four-dimensional Einstein-Maxwell theory. The NP variables are given by
\begin{align}
&\rho=-\frac{1}{\varrho},\quad \varrho=r+i\Sigma,\quad \alpha=\frac{\alpha^0}{\varrho},\quad  \alpha^0=\frac{\bz}{2\sqrt2}, \quad  \beta=-\xbar\alpha ,\nn \\
& \tau=0=\lambda,\quad \mu=-\frac{1}{2\xbar\varrho}-\frac{r\Psi_2^0+\phi_1^0 \xbar\phi_1^0}{\varrho^2\xbar\varrho}, \quad  \gamma=-\frac{\Psi_2^0}{2\varrho^2}-\frac{\phi_1^0\xbar\phi_1^0}{\varrho^2\xbar\varrho},\nn \\
& \Psi_2=\frac{\Psi_2^0}{\varrho^3} + \frac{2\phi_1^0 \xbar \phi_1^0}{\varrho^3 \xbar\varrho },  \quad \nu = \frac{\xbar\eth \Psi_2^0}{2\varrho^2} - \frac{ i \Psi_2^0  \xbar\eth \Sigma}{\varrho^3} + \frac{\xbar\phi_1^0 \xbar\eth\phi_1^0}{\varrho^2\xbar\varrho} - \frac{2i \phi_1^0 \xbar\phi_1^0 \xbar\eth \Sigma}{\varrho^3\xbar\varrho}, \nn \\
&\phi_1=\frac{\phi_1^0}{\varrho^2},\quad\quad  \phi_2= - \frac{\xbar\eth \phi_1^0}{\varrho^2} + \frac{2 i \phi_1^0 \xbar\eth \Sigma }{\varrho^3}, \\
&\Psi_3=-\frac{\xbar\eth \Psi_2^0}{\varrho^3} + \frac{3 i\Psi_2^0 \xbar\eth \Sigma}{\varrho^4} - \frac{2\xbar\phi_1^0 \xbar\eth\phi_1^0}{\varrho^3\xbar\varrho}+ \frac{6i \phi_1^0 \xbar\phi_1^0 \xbar\eth \Sigma}{\varrho^4\xbar\varrho},\nn \\
&\Psi_4= \frac{\xbar\eth^2 \Psi_2^0}{2\varrho^3} - \frac{3i\xbar\eth\Psi_2^0 \xbar\eth \Sigma + i \Psi_2^0 \xbar\eth^2 \Sigma }{\varrho^4} - \frac{6\Psi_2^0  (\xbar\eth \Sigma)^2}{\varrho^5}  + \frac{\xbar\phi_1^0 \xbar\eth^2 \phi_1^0}{\varrho^3\xbar\varrho} \nn \\
&\qquad \qquad - \frac{6i\xbar\phi_1^0 \xbar\eth\phi_1^0\xbar\eth \Sigma + 2i\xbar\phi_1^0  \phi_1^0 \xbar\eth^2 \Sigma }{\varrho^4\xbar\varrho} - \frac{12\xbar\phi_1^0 \phi_1^0 (\xbar\eth \Sigma)^2}{{\varrho^5\xbar\varrho}}.\nn
\end{align}
Clearly, the conditions $\bp\Psi_2^0=0$ and $\bp\phi_1^0=0$ imply that $\Psi_2^0$ and $\phi_1^0$ are holomorphic functions which we denote as $\Psi_2^0=L(z)$ and $\phi_1^0=Q(z)$. Then, the twist function $\Sigma$ is determined from the differential equation
\be
\Sigma + \frac{(1+z\bz)^2}{2} \p\bp \Sigma=\frac{i}{2}(L-\bL).
\ee
Now the above equation is the only constraint left for the whole system, which is the same as pure Einstein theory.

\bibliography{ref}

\end{document}